\documentclass[
  journal=largetwo,
  manuscript=research-article,
  year=2024,
  volume=37,
]{cup-journal}

\usepackage{amsmath}
\usepackage[nopatch]{microtype}
\usepackage{booktabs}
\usepackage{multirow}
\usepackage{longtable}
\usepackage{subcaption}
\usepackage{fancyhdr}
\usepackage{xcolor}

\usepackage{graphicx}

\DeclareGraphicsExtensions{.eps, .pdf}

\title{Orbit determination of visual binary systems observed with CST telescope in 2010-2013}

\author{F. M. Rica}
\affiliation{Federaci\'on Extremeña de Astronom\'ia, C/Jos\'e Ru\'iz Azor\'in, 14, 4º D, M\'erida 06800,  Spain}
\email[F. M. Rica]{frica0@gmail.com}

\author{R. Barrena}
\affiliation{Instituto de Astrofísica de Canarias, C/ Vía Láctea s/n, La Laguna 38205, Spain}
\alsoaffiliation{Universidad de La Laguna, Facultad de Ciencias, Departamento de Astrofísica, Avenida Astrofísico Francisco Sánchez, s/n, 38206, La Laguna, Tenerife, Spain}

\author{J. A. Henríquez}
\affiliation{Grupo de Observadores Astronómicos de Tenerife (GOAT), Spain}
\alsoaffiliation{Astronomical Observatory MPC J-48, C/San Pancracio, no 26, Urb. Mackay, San Cristóbal de la Laguna E-38205, Spain}

\author{G. Vázquez}
\affiliation{Astronomical Observatory MPC J-48, C/San Pancracio, no 26, Urb. Mackay, San Cristóbal de la Laguna E-38205, Spain}
\alsoaffiliation{Agupación Astronómica de Tenerife, post office box 10644-38080, Tenerife, Spain}

\author{C. Vázquez}
\affiliation{Astronomical Observatory MPC J-48, C/San Pancracio, no 26, Urb. Mackay, San Cristóbal de la Laguna E-38205, Spain}
\alsoaffiliation{Agupación Astronómica de Tenerife, post office box 10644-38080, Tenerife, Spain}

\author{D. Hernández}
\affiliation{Sociedad para el Estudio y el Conocimiento de la Astronomía (SECAT), Tenerife, Spain}

\author{J. L. del Rosario}
\affiliation{Sociedad para el Estudio y el Conocimiento de la Astronomía (SECAT), Tenerife, Spain}

\keywords{visual binary stars; visual binary stars; orbit determination; fundamental parameters of stars } 

\begin{document}

\begin{abstract}
We present new orbital solutions for 15 binaries, which were astrometrically measured by our team during 2010-2013, using the FastCam ``lucky-imaging'' camera installed at the 1.5-m Carlos S\'anchez Telescope (CST) at the Observatorio del Teide, Tenerife (Spain). We present first orbital solutions for BU 1292, STF 147, HDS 1898 and STT 325 and revise orbital solutions for AG 14, D 5 AB, A 1581, HO 525 AB, WOR 19, A 1999, HU 572, HU 742, COU 227, BU 696 AB, and A 893. We apply two orbital calculation techniques, the ``three-dimensional grid search method'', first described by Hartkopf, McAlister, \& Franz (1989), and the Docobo's analytical method (Docobo 1985).
We use our tool ``Binary Deblending'', based on deblending the entire observed multiband photometry into fundamental and photometric parameters for each stellar component based on PARSEC isochrones. We also obatain the total mass for all systems. Our findings include the identification of a binary system consisting of two M-type dwarfs (WOR 19), a binary of evolved components (twin F6IV-V stars) in BU 1292, accompanied by a newly discovered wide (10.5") and faint companion with \emph{G} = 17.05 mag. Additionally, we explore the X-ray emission system STF 147 and a very young quadruple system, WDS 04573+5345. This comprehensive analysis significantly contributes to our understanding of the formation and evolution of stellar systems.

\end{abstract}

\section{Introduction}

One of the relevant goals in stellar astrophysics is understanding of formation and evolution of the stars. When these stars form as part of binary and multiple systems, it is possible to determine the fundamental stellar parameters. The most important feature is the stellar mass, which can be determined independently of the models through the analysis of orbital solutions. A particular subsample of binary and multiple systems are visual systems that consist of stars visually resolved using optical telescopes. Analysing the photometry (combined or individual) of these systems and the trigonometric parallax, it is possible to estimate the fundamental parameters for the stellar component. In addition, if an orbital solution is known,  we can determine the sum of mass or even the individual masses, whether additional information like a spectroscopic orbital solution is available (Abushattal et al. 2020), so calibrating the stellar models used. 

Currently, several teams worldwide are engaged in studying visual binary systems and calculating their orbital solutions. Important contributions in this sense are those of Horch et al. (2021), Mendez, Clavería \& Costa (2021), and Tokovinin (2021).

This paper is a continuation of the visual binary star observation project carried out in the 1.5-m Carlos S\'anchez Telescope (hereafter CST) at Teide Observatory, using the Fastcam lucky imaging camera. A first work for this project was published by Rica et al. (2012c). In addition, we have recently published 886 new astrometric measures of 447 visual binaries observed in 2010-2013 using the lucky-imaging technique (see Rica et al. 2022, hereafter Paper I). We measured visual binaries (showing angular separation between 0.13 'and 17.17') with no astrometric point in the last 10-15 years, some of them with no measurements for even more than 100 years. Some of the binaries showed important residuals with respect to their previously published orbital solutions. In this work, we improve some published orbits and report new orbits for some of them. We also present an astrophysical study using multiband photometric data and trigonometric parallax from the literature to obtain the photometry and fundamental parameters for each stellar component.

This paper is organized as follows. In Section 2, we detail the methods used in the orbital calculation. In Section 3, we list and describe the visual binaries for the orbits obtained. We provide a detailed description of the figures and data tables that illustrate the orbital solutions, along with the astrophysical and dynamical properties. In Section 4, we expose our analysis. In Section 5, we detail our results for each binary system. Finally, the summary and conclusions are presented in Section 6.

\section{Methods of orbital calculation}

The orbits present in this work are computed by employing two calculation methods:

\begin{enumerate}

\item For the binaries A 1581, WOR 19, HU 572, HU 742, STT 325 and COU 227 we employ Docobo's apply method (Docobo 1985). This method is summarized in Docobo et al. (2000) and Tamazian et al. (2002). It can be used even when the observed orbital arc is relatively short and linear. This method determines the set of Keplerian orbits whose apparent orbits pass through three base points. The selection of the best orbital solution is based on different criteria. We use the root mean square (RMS) residual on position angle and angular distance. As a final step, the orbits are improved using the differential correction method, developed by Heintz (1978), in its polar form. 

\item The ``three-dimensional grid search method'', was first proposed by Hartkopf, McAlister, \& Franz (1989) and further modified by Mason, Douglass, \& Hartkopf (1999). This method starts from three known elements: the orbital period (\emph{P}), the epoch of periastron (\emph{T}) and the eccentricity (\emph{e}). These parameters are used to obtain the geometric elements \emph{a} (semimajor axis), \emph{i} (inclination), \emph{$\Omega$} (ascending node) and \emph{$\omega$} (argument of periastron) by the method of least squares. In practice, we define a space of possible values for \emph{P}, \emph{T} and \emph{e} for which we determine a possible orbital solution. The solution with the smallest residuals is chosen. Finally, we apply a least square refinement, if it was necessary, using the formula for differential corrections in rectangular coordinates (Heintz 1967a). This technique can also be used in case the observed orbit arc is relatively short and closely linear. Therefore, due to the intrinsic features and data available, the orbits we obtain with this method are AG 14, BU 1292, STF 147, D 5 AB, HO 525 AB, A 1999, HDS 1898, BU 696 AB and A 893.
\end{enumerate}

These two methods are choosen because their final orbital solutions are robust and they are easy to implement and use. For some orbital solutions, we list the errors of the orbital elements (if an orbital element has no errors reported, it is due to the impossibility to constrain it). We obtain formal errors at $1\sigma$ uncertainty interval for the orbital elements using the covariance matrix for A 1581, WOR 19, HU 572 and BU 696. When the orbit is poorly determined, we consider a more realistic error calculation method. This is based on the distribution of the orbital elements for the orbital solutions that produce a difference of one with respect to the minimum reduced-$\chi^2$. This is so applied in the cases of AG 14, BU 1292, A 1999, HDS 1898 and A 893. 

The weights for astrometric measures are assigned using the data-weighting scheme published in Rica et al. (2012c), which is function of different criteria: the observational method, the aperture of the telescope, the experience of the observer, the number of nights observed. The initial \emph{$\theta$} weights are four times larger than \emph{$\rho$} weights (Heintz 1978a) for visual measures. The astrometric points with residuals larger than $3\sigma$ for each group (selected mainly as function of the observational technique) are assigned as zero weight. Later, the non-zero weight measures are reassigned following the work of Irwin et al. (1996). 

\section{Orbital solutions}

Many of the binaries in Paper I have already published orbital parameters. However, our measures for some of these binaries exhibit significant deviations from the orbital solution ephemerids. Therefore, if our measures agree with recent historical measures, we consider that the binary needs an improved orbital solution.

A large amount of visual binary systems have orbital periods spanning one or more centuries, making the determination of their definitive orbits a multigenerational endeavor. It is not unusual to see multiple orbital solutions published, each of them introducing significant improvements over time.

In our study, we calculate improved orbital solutions for WDS 01055+2107 (AG 14), WDS 04573+5345 (D 5 AB), WDS 08031-0625 (A 1581), WDS 08231+2001 (HO 525 AB), WDS 08286+3502 (WOR 19), WDS 12154+4008 (A 1999), WDS 13091+2127 (HU 572), WDS 14087+3341  (HU 742), WDS 21115+2144 (COU 227), WDS 22045+1551 (BU 696 AB) and WDS 22052+2952 (A 893). In Section 5, we provide details on the specific aspects that are improved. On the other hand, we have found no orbital solutions published previously for other systems and have computed a first orbital solution for them. This is the case of WDS 01078+0425 (BU 1292), WDS 01417-1119 (STF 147), WDS 13327+2230 (HDS 1898), and WDS 17130+0745  (STT 325).

Before using an orbital calculation method, we correct the position angles, \emph{$\theta$}, for precessions to 2000 equinox. Then, we plot \emph{$\theta$} and \emph{$\rho$} against time to detect outliers astrometric measures or quadrant reversal problems. Outliers positions are rejected for the orbital solution calculation assigning zero weight.

Most of these orbital solutions were published in different circulars of IAU Commission G1 (Binary and Multiple Star Systems) bewteen 2012 and 2014. The references to these circulars are given within the analysis for each binary system (see Section 5). In this paper, we publish the complete orbital parameters with their errors and weighted mean residuals, orbital plots and the astrophysical analysis for the first time. 

Table 1 lists, for each binary, the final orbital elements and root mean square (RMS). These are, in both cases, the weighted averages obtained using the data-weighting scheme described in this Section. Table 2 lists preliminary orbital solutions but without error estimates. For these binary stars, the astrometric data cover only a very short arc of the orbital path, significantly limiting the constraints on the orbits. Additionally, the quality of the astrometric measurements is relatively low, further reducing the reliability of the derived parameters. As a result, it is not possible to estimate formal uncertainties in the orbital elements for this objects.

Table 3 lists stellar data and is organized as follows: the WDS and the Hipparcos numbers in columns (1) and  (2), respectively; the WDS spectral type (hereafter SpT) and the \emph{V} apparent magnitude of the A and B components in columns (3)-(5); the dynamical parallax and the SpT determined in this work are listed in columns (6)-(7); in columns (8) and (9) the Hipparcos trigonometric parallax (van Leeuwen 2007) and the total mass of the binary (with the estimated standard errors), respectively, as calculated from the Hipparcos parallax. In columns (10) and (11), the Gaia DR3 parallax and the total mass of the binary. For WDS 08031-0625 Gaia parallaxes are listed for both stellar components.

\begin{table*}[t!]
\caption{Orbital parameters, parallaxes and residuals. }
\label{tab_1}
\begin{tabular}{llllll}
\midrule
Name & AG   14     & BU 1292      &  A 1581  & WOR 19      & A 1999  \\
WDS  & 01055+2107  & 01078+0425   & 08031-0625  & 08286+3502  & 12154+4008 \\
ADS  & 896         & 929          &  6547   &             & 8480    \\
HIP  & 5115        & 5297         &  39383   & 41554       & 59776  \\
\midrule
P (years)   &  $498.66_{-252}^{+1806}$  &  $285.28_{-74}^{+42}$  &   $1225_{-809}^{+775}$    & 50.138 ± 0.22  & $200.01_{-82}^{+478}$ \\
T (BY)      &  $2465.42_{-253}^{+1805}$  &  $2020.73_{-141}^{+179}$   &   $3121_{-798}^{+799}$  & 2000.129 ± 0.31     &  $2014.08_{-10.5}^{+4.25}$  \\
e           &  $0.70_{-0.20}^{+0.20}$  &  $0.214_{-0.21}^{+0.09}$   &   $0.58_{-0.08}^{+0.32}$  & 0.184 ± 0.005  &  $0.589_{-0.09}^{+0.18}$ \\
a (arcsec)  &  $1.06_{-0.33}^{+1.93}$  &  $0.309_{-0.057}^{+0.039}$   & $1.98_{-0.96}^{+0.92}$   & 0.604 ± 0.010  &  $0.983_{-0.356}^{+1.485}$ \\
i (degs)    &  $159.0_{-16.8}^{+14.0}$  &  $139.2_{-12.0}^{+22.0}$   &  $69.1_{-31.0}^{+6.3}$  & 136.0 ± 1.7    & $58.0_{-22.8}^{+11.2}$ \\
$\omega$ (degs)    & 190.9   & 194.8   &  63.6    & 123.7 ± 3.5   & $279.4_{-43.3}^{+44.3}$ \\
$\Omega$ (degs)    & 265.5   & 52.9    &  124.9  & 63.0 ± 2.6   &  $198.1_{-10.9}^{+32.1}$ \\
\hline
Residuals:         &         &         &      &      &      \\
\hline
RMS($\theta$)(degs)     & 3.85    & 5.89      &  1.70    & 2.60    & 12.97   \\
RMS($\rho$) (arcsec)    & 0.038   & 0.021     &  0.116   & 0.040      & 0.026  \\
\midrule
\hline
Name    & HU 572   & HDS 1898  & BU 696 AB & A 893 & \\
WDS     & 13091+2127 & 13327+2230  & 22045+1551 & 22052+2952 &\\
ADS     & 8799    &     & 15599  & 15610 & \\
HIP      & 64175    & 66072  & 108961 & 109018  &	\\
\midrule
P (years)         & 102.898 ± 0.92   &  $30.87_{-1.80}^{+2.40}$  & 160.3 ± 7.3  & $105.9_{-6.0}^{+8.7}$ & \\
T (BY)       & 1986.222 ± 0.20   &  $1998.89_{-3.25}^{+2.81}$  & 2006.74 ± 1.00  & $1990.8_{-3.4}^{+3.0}$  & \\
e               & 0.516 ± 0.008   &  $0.718_{-0.04}^{+0.02}$ &  0.909  ± 0.007  & $0.44_{-0.05}^{+0.03}$ & \\
a (arcsec)   & 0.400 ± 0.005  &  $0.24_{-0.02}^{+0.02}$   &  0.386 ± 0.035 & $0.225_{-0.011}^{+0.023}$ &  \\
i (degs)        & 145.0 ± 1.9     &  $67.8_{-2.6}^{+1.7}$  &  84.9 ± 0.8 & $148.1_{-11.4}^{+6.9}$ & \\
$\omega$ (degs)      & 30.9 ± 4.4   &  $93.0_{-3.9}^{+2.9}$  & 46.4 ± 5.4  & $190.4_{-23.5}^{+26.9}$ & \\
$\Omega$ (degs)    & 153.7 ± 4.0   &  $238.3_{-13.6}^{+13.1}$  &  172.2 ± 0.5 & $195.5_{-15.8}^{+20.0}$  & \\
\hline
Residuals:         &             &    &   &  & \\
\hline
RMS($\theta$)(degs)     & 4.18  & 1.84   &  2.98  & 2.51  & \\
RMS($\rho$) (arcsec)     & 0.029   & 0.003 & 0.018  & 0.014 & \\
\midrule
\hline
\end{tabular}
\end{table*}

\begin{table*}[t!]
\caption{Preliminary Orbital parameters, parallaxes and residuals.} 
\label{tab_2}
\begin{tabular}{lllllll}
\midrule
Name & STF 147    & D 5 AB   &   HO 525 AB  & HU 742      &  STT 325  & COU 227 \\
WDS  & 01417-1119 & 04573+5345 & 08231+2001  & 14087+3341  &  17130+0745 & 21115+2144 \\
ADS  & 1339       & 3536   &   6776   & 9126        &  10398  &    \\
HIP  &  7916       & 23040   &   41098 & 69102       &  84230   & 104612 \\
\midrule
P (years)  & 627.371     & 824.37 &  935.00  & 726.197    & 2838.434   & 198.657 \\
T (BY)      & 2014.730    & 2173.12  &  1967.00  & 1984.006    & 2130.180  & 2031.701 \\
e           & 0.919       & 0.35   &  0.700  & 0.643     & 0.662  & 0.299 \\
a (arcsec)  & 2.499       & 1.558   &  0.844  & 1.457     & 3.344 & 0.669   \\
i (degs)    & 81.98       & 106.2    &  39.1   & 107.74       & 77.91  & 52.85  \\
$\omega$ (degs)     & 174.7    & 34.9 &   115.7  & 38.1     & 348.8  & 44.5 \\
$\Omega$ (degs)    & 86.9     & 153.3 & 144.9  & 118.9    & 6.2  & 126.3 \\
\hline
Residuals:         &         &         &      &      &    &   \\
\hline
RMS($\theta$)(degs)       & 7.92      & 0.88  &   3.92  & 4.73       & 2.91   & 1.44  \\
RMS($\rho$) (arcsec)      & 0.018     & 0.012   &  0.025 & 0.028     & 0.057 & 0.021   \\
\midrule
\hline
\end{tabular}
\end{table*}

Table 4 shows the fundamental parameters for each stellar component determined in this study. In columns (1) and (2), we show the WDS identificator and the star name; in columns (3), the absolute magnitudes in V-band and, in the following columns, the \emph{T}$_{eff}$, log \emph{g}, stellar mass, and SpT are reported. 

The dynamical parallaxes are estimated using the well-known Baize-Roman\'i method (Baize \& Roman\'i 1946), that was improved by Heintz (1978) and by Docobo \& Andrade (2013), updating the mass-luminosity relationship. This Baize \& Roman\'i method, from the 3rd Kepler law, computes the parallax and the individual stellar masses using the orbital period, the semimajor axis and the bolometric apparent magnitudes for the components. A mass-luminosity relation is used for stars of V and IV luminosity class. This technique allows us to check the stellar masses and parallaxes from the literature or computed by us.

Figures 2-4 show all the orbit plots determined in this paper. The black solid thick ellipse is our solution and the grey thin ellipse is the previously calculated orbit in the literature. North is down and East is right. The filled black circle is the main stellar component at (0,0) coordinate. Astrometric measures are represented by green ``+'' (micrometric measures), red square are the Hipparcos and the Tycho-2 measures and large blue circle (speckle). Small blue circle are photographic and CCD measures, pink ``+'' (our measures published in Rica et al. 2022). Rejected measures are plotted with red ``X''.
\begin{center}
\begin{table*}[t]
\caption{Stellar Data}
\label{Table3}
\begin{tabular}{ll m{1cm} m{0.9cm} m{0.9cm} m{1cm}  m{2cm} llll}
\multicolumn{2}{l}{Star} & WDS  & \multicolumn{4}{c}{Present Work}  & \multicolumn{2}{c}{Hipparcos} & \multicolumn{2}{c}{Gaia DR3} \\
\hline
WDS & HIP      & Spectral Type & $V_A$ (mag) & $V_B$ (mag) & Dyn. $\pi$ (mas) & Spectral Type              & $\pi$ (mas)   & $\sum$ ($M_\odot$) & $\pi$ (mas)   & $\sum$ ($M_\odot$)   \\
\hline
01055+2107    & 5115     & K0            & 9.73        & 10.23       & 14.6             & G7V+K0V                    & 10.02 ± 2.56  & 4.8                & 15.09 ± 0.04  & 1.40                 \\
01078+0425    & 5297     & F5            & 9.8         & 10.3        & 5.2              & F6IV-V+F5V                 & 3.11 ± 1.52   & 12.1               & …             & …                    \\
01417-1119    & 7916     & F5V+F7V       & 6.1         & 7.2         & 24.5             & F4IV-V+F7V                 & 25.35 ± 0.68  & 2.4                & …             & …                    \\
04573+5345    & 23040    & A1V           & 4.48        & 7.73        & …                & (B9V-IV+F5/6V) + F0V + K0V & 8.77 ± 0.70   & 4.5 ± 1.1          & 9.44 ± 0.02   & 3.6 ± 0.1                                                 \\
08031-0625    & 39383    & K0            & 9.88        & 10.18       & 14.7             & (G9V+K2V) + G8V            & 16.94 ± 2.21  & 1.1                & \begin{tabular}[c]{@{}l@{}}14.48 ± 0.03\\ 14.64 ± 0.04\end{tabular} & \begin{tabular}[c]{@{}l@{}}1.70\\ 1.65\end{tabular} \\
08231+2001    & 41098    & F5            & 9.73        & 9.78        & 6.4              & F4V+F4V                    & 7.15 ± 1.34   & 1.9 ± 0.6          & 7.09 ± 0.23   & 1.9 ± 0.2            \\
08286+3502    & 41554    & M0V           & 11.45       & 11.58       & 45.5             & M1.5V+M1.5V                & 50.82 ± 6.27  & 0.67 ± 0.04        & …             & …                    \\
12154+4008    & 59775    & G9IV          & 8.96        & 10.98       & 27.4             & K1.5V+K6V                  & 24.37 ± 1.35  & 1.6  ±   2.3       & 25.14 ± 0.25  & 1.5  ±   2.0         \\
13091+2127    & 64175    & G5            & 8.69        & 9.68        & 14.7             & G0V+G8V                    & 12.75 ± 1.43  & 2.9 ± 1.0          & 8.79 ± 0.30   & 8.9 ± 1.0            \\
13327+2230    & 66072    & …             & 10.17       & 10.70       & 24.2             & K3V+K4.5V                  & 21.71 ± 1.64  & 1.44 ± 0.52        & …             & …                    \\
14087+3341    & 69102    & K0III         & 9.00        & 10.40       & 15.2             & G4V+K1V                    & 13.26 ± 1.37  & 2.5                & 13.43 ± 0.54  & 2.4                  \\
17130+0745    & 84230    & F0            & 7.15        & 8.94        & 11.7             & (A7V+G9V) + F6V            & 10.46 ± 0.70  & 4.1                & 8.61 ± 0.65   & 7.3                  \\
21115+2144    & 104612   & K2            & 10.01       & 11.58       & 18.5             & K0.5V+K5V                  & 13.06 ± 1.48  & 3.4                & 13.90 ± 0.39  & 2.8                  \\
22045+1551    & 108961   & G0V           & 8.50        & 8.80        & 9.27             & F6V+F7V                    & 10.71 ± 0.90  & 1.8 ± 0.7          & …             & …                    \\
22052+2952    & 109018   & F8V           & 9.08        & 10.08       & 7.50             & F5V+G1V                    & 7.77 ± 1.08   & 2.2 ± 1.1          & …             & …                    \\
\hline
\end{tabular}
\end{table*}
\end{center}

\begin{center}
\begin{table*}[t]
\caption{Fundamental stellar data}
\label{Table_4}
\begin{tabular}{llllllll}
\hline
Component      & ID            & Mv    & \emph{T}$_{eff}$ (K) & log \emph{g} & Mass ($M_\odot$) & Radius ($R_\odot$) & SpT    \\
\hline
01055+2107  A  & HD 6439  A    & 5.55  & 5564     & 4.60  & 0.89      & 0.74        & G7V    \\
01055+2107  B  & HD 6439  B    & 6.05  & 5251     & 4.63  & 0.82      & 0.68        & K0V    \\
01078+0425  A  & HD 6698 A     & 3.07  & 6389     & 4.04  & 1.27      & 1.69        & F6IV-V \\
01078+0425  B  & HD 6698 B     & 3.57  & 6489     & 4.23  & 1.16      & 1.30        & F5V    \\
01417-1119  A  & HD 10453 A    & 3.10  & 6576     & 4.07  & 1.20      & 1.59        & F4IV-V \\
01417-1119  B  & HD 10453 B    & 4.23  & 6263     & 4.38  & 1.03      & 1.03        & F7V    \\
04573+5345 Aa  & 7 Cam Aa      & -0.77 & 11750    & 3.74  & 3.61      & 4.23        & B9IV-V \\
04573+5345 Ab  & 7 Cam Ab      & 3.76  & 6640     & 4.36  & 1.30      & 1.25        & F5/6V  \\
04573+5345 B   & 7 Cam B       & 2.56  & 8200     & 4.31  & 1.70      & 1.51        & F0V    \\
04573+5345 C   & 7 Cam C       & 5.77  & 5400     & 4.59  & 0.91      & 0.80        & K0V    \\
08031-0625  Aa & HIP 39383  Aa & 6.15  & 5264     & 4.62  & 0.75      & 0.60        & G9V    \\
08031-0625  Ab & HIP 39383  Ab & 6.78  & 4886     & 4.65  & 0.69      & 0.48        & K2V    \\
08031-0625  B  & HIP 39383  B  & 5.94  & 5400     & 4.61  & 0.77      & 0.67        & G8V    \\
08231+2001 A   & HIP 41098 A   & 3.93  & 6628     & 4.40  & 1.17      & 1.07        & F4V    \\
08231+2001 B   & HIP 41098 B   & 3.98  & 6592     & 4.41  & 1.16      & 1.05        & F4V    \\
08286+3502  A  & GJ 308 A      & 9.96  & 3665     & 4.84  & 0.43      & 0.39        & M1.5V  \\
08286+3502  B  & GJ 308 B      & 10.09 & 3648     & 4.85  & 0.42      & 0.38        & M1.5V  \\
12154+4008 A   & HD 106592  A  & 5.88  & 5180     & 4.52  & 0.81      & 0.78        & K1.5V  \\
12154+4008 A   & HD 106592  B  & 7.90  & 4233     & 4.62  & 0.66      & 0.62        & K6V    \\
13091+2127  A  & HD 114255  A  & 4.20  & 6062     & 4.30  & 1.04      & 1.13        & G0V    \\
13091+2127  B  & HD 114255  B  & 5.19  & 5626     & 4.49  & 0.91      & 0.85        & G8V    \\
13327+2230 A   & HIP 66072 A   & 6.84  & 4824     & 4.62  & 0.67      & 0.62        & K3V    \\
13327+2230 B   & HIP 66072 B   & 7.37  & 4552     & 4.64  & 0.63      & 0.59        & K4.5V  \\
17130+0745  Aa & HD 155714 Aa  & 2.03  & 7506     & 4.06  & 1.77      & 2.05        & A7V    \\
17130+0745  Ab & HD 155714 Ab  & 5.53  & 5374     & 4.56  & 0.95      & 0.84        & G9V    \\
17130+0745  B  & HD 155714 B   & 3.71  & 6439     & 4.32  & 1.30      & 1.23        & F6V    \\
21115+2144 A   & COU 227 A     & 5.55  & 5300     & 4.49  & 0.84      & 0.82        & K0.5V  \\
21115+2144 B   & COU 227 B     & 7.12  & 4514     & 4.60  & 0.70      & 0.65        & K5V    \\
22045+1551  A  & BU 696 A      & 3.62  & 6250     & 4.22  & 1.30      & 1.39        & F6V    \\
22045+1551  B  & BU 696 B      & 3.83  & 6180     & 4.27  & 1.25      & 1.28        & F7V    \\
22052+2952 A   & A   893 A     & 3.45  & 6600     & 4.26  & 1.32      & 1.32        & F5V    \\
22052+2952 B   & A   893 B     & 4.45  & 6064     & 4.43  & 1.11      & 1.00        & G1V   \\
\hline
\end{tabular}
\end{table*}
\end{center}

\section{Determining fundamental parameters and other astrophysical properties} 

For each binary studied in this work, we conduct an astrophysical analyse using the multiband photometric data, the trigonometric parallax and the \emph{V} differential magnitude between both stellar components obtained from the literature. Our study involves the use of evolutionary isochrones to separate the combined observed \emph{UBVIJHK} photometry of the binary systems (primarily sourced from the Hipparcos, Tycho-2 and 2MASS catalogues mainly) into synthetic photometry and fundamental parameters for each stellar component. To obtain the required isochrones, we use the CMD 3.3 evolutionary isochrones \footnote[1]{CMD is a service mantained at the Osservatorio Astronomico di Padova, composed by a set of routines that provide interpolated isochrones in a grid, together with derivatives such as luminosity functions, simulated star clusters, etc. The photometry can be produced for many different broad- and intermediate-band systems, including non-standard ones. Online tool: \url{http://stev.oapd.inaf.it/cgi-bin/cmd}.} online tool based on PARSEC release v1.2S + COLIBRI \verb|S_35| (Bressan et al. 2012, Chen et al. 2014, 2015, Tang et al. 2014, Marigo et al. 2017, Pastorelli et al. 2019).

We also use the ``Binary Deblending'' tool, version v5.0, that was designed and developed by our team. This tool takes the following data as input: 

\begin{itemize}
	 \item the unreddening combined photometry in \emph{UBVIJHK} bands; 
	 \item the differential magnitude in V-band (\emph{$\Delta$V}) for the stellar components;
	 \item the trigonometric parallax of the system;
	 \item and the \emph{UBVIJHK} synthetic photometry from PARSEC isochrones, covering a wide range of ages and  metallicities [Fe/H].
\end{itemize}

This tool searches for two entries in the evolutionary isochrones, aiming to minimize the $\chi^2$ between the combined observational photometry and that obtained from PARSEC model. Our tool returns the synthetic photometry in \emph{UBVIJHK} bands, SpT, and fundamental parameters (masses, \emph{T}$_{eff}$, log \emph{g}, luminosity and radius) for each component. 

The $\chi^2$ is calculated as 

\begin{equation}
\label{eq:first}
\begin{aligned}
 \chi^2 = \sum_{i}{\frac{Mo_i - Mm_i}{\sigma_i}^2 }
\end{aligned}
\end{equation}

where $Mo_i$ and $Mm_i$ are the observed and modelled combined magnitudes for each photometric band, $\sigma_i$ is the observational photometric error reported in literature for each filter \emph{UBVIJHK}. This method shows the minimum $\chi^2$ solution for each metallicity and age. We can choose the solution as that with the minimum $\chi^2$,  or alternatively other different one. This procedure provides a comparative table between the combined observed photometry and the combined model one. In addition, we can also get information about the fundamental parameters for the components of the binary system. In some cases, it is not possible estimate the error interval for the metallicity and age of the stellar components, in particular when they are not evolved stars.

The reddening in the line of sight is estimated using the maps of Schlafly and Finkbeiner (2011). The resulting values are scaled to the Hipparcos or Gaia distance using the formula published by Anthony-Twarog \& Twarog (1994). In addition, we also use the Stilim web that produces tridimensional maps of the local InterStellar Matter (ISM). These maps are based on measurements of starlight absorption by dust (reddening effects) or gaseous species (absorption lines or bands) and is based on the inversion of reddening estimates towards 71,000 target stars.

In this paper, we estimate the approximate stellar age from the galactocentric velocity \emph{U}, \emph{V}, \emph{W} (Przybylski 1962) and using the Eggen diagrams (see Figs. 1 in Eggen (1969a) and Eggen (1969b)). These diagrammes provide information relative to the stellar age, by distinguishing between young and old stars. In addition to these diagrammes, we used the kinematic age parameter of Grenon (1987), fG. Bartkevicius \& Gudas (2002) determined the relation between fG and the age in order to distinguish between different age groups.  

Statistically they found that stars with $fG < 0.20$ belong to the young-middle age group (with an age less than $3-4$ Gyr) of the thin disk population, while the stars with $0.20 < fG < 0.35$ belong to the old (with age of $3-10$ Gyr) thin disk population. The stars with $0.35 < fG < 0.70$ belong to the thick disk population (age greater than 10 Gyr), and the stars with $fG > 0.70$ belong to the halo population. 

When a target shows evidences to be a young system, we use the BANYAN $\Sigma$ online tool  (Gagn\'e et al. 2018) to check the membership to any of the 27 known and well-characterized young associations within 150 pc with ages between $1-800$ Myr. The algorithm of this tool is based on the comparison between the galactic position \emph{X},\emph{Y},\emph{Z} and the space velocity \emph{U}, \emph{V}, \emph{W} using a Bayesian classifier.

\section{Discussion on individual binary stars} 
 
\underline{WDS 01055+2107 = AG 14 = ADS 896} \\

The system AG 14 ( = HD 6439 = HIP 5115) is composed of stars with \emph{V} magnitudes of 9.7 and 10.2 separated by 0.8”, with a previous orbit obtained by Heintz (1998). The magnitudes listed in the WDS main index are not correct. Based in the Hipparcos photometry and the magnitudes estimation listed in the WDS historical astrometric measures, we estimate a $\Delta V = 0.5$ mag.

The first measure was performed in 1901.92 (Hussey 1903) and this binary star has 65 astrometric measures that cover an arc of about 262 degrees. Our measures conducted in 2011-2012 (Paper I) revealed that the orbit is opening up compared to that calculated by Heintz. Our new orbital solution was previously announced in the IAUDS No. 184 (Rica 2014a) and has an orbital period about three times longer than the previous orbit.

The Henry Draper Catalogue and Extension (Cannon \& Pickering 1918-1924) classifies this object as a star with a SpT of K0. Stephenson \& Sanwal (1969) determined individual SpT of K0+K2. Using our tool ``Binary Deblending v5.0'' and the Gaia DR3 parallax, we estimate SpT of G7V+K0V, with a total mass of 1.7 $M_\odot$.

Our dynamical parallax, $14.6_{-0.1}^{+1.5}$ mas, is in good agreement with the Gaia DR3 parallax (15.09 ± 0.04 mas). The Hipparcos parallax, $10.55 \pm 0.89$ mas, gives excessive total mass. On the other hand, our dynamical total mass, $1.54_{-0.08}^{+0.02}$ $M_\odot$, is within the expected value obtained in our analysis. Gaia DR3 lists both stellar components (but no astrometric solution is provided for the secondary component) with a RUWE of 1.9 for the primary star. This high value indicates an issue with the single-star astrometric model and it is likely caused by the presence of the close secondary.

In the literature we find radial velocity with coherent values of about -3 km s$^{-1}$ (Griffin 1986; Tokovinin \& Smekhov 2002; Gontcharov 2006). Here, we present for the first time the galactocentric velocity (\emph{U}, \emph{V}, \emph{W}) of $(+6.3, -6.0, -4.8)$ km s$^{-1}$, which has not been previously reported in the literature. This velocity corresponds to that of stars belonging to the young stellar population in the thin Galactic disk. The kinematic of this binary system has not been identified with any known young kinematic group. \\

\underline{WDS 01078+0425 = BU 1292 = ADS 929} \\

BU 1292 ( = HD 6698 = HIP 5297) was discovered by Burnham (1903) at the Yerkes Observatory in 1901. This binary system is composed of stars separated by 0.3”. None previous orbit is listed in the Sixth Orbit Catalog. Therefore, the orbital solution presented in this study is the first one obtained for BU 1292. Our final orbit shows a low eccentricity and an orbital period of approximately 285 years. The calculated periastron happened near 2021 and the last astrometric measure listed in the WDS database was obtained in 2013, indicating the need for continuous monitoring of this binary system \footnote[2]{In the following years the angular separation will be about 0.23-0.24” and therefore accesible to small aperture telescopes}. Additionally, the small Hipparcos parallax ($3.11 \pm 1.52$ mas) is likely affected by systematic errors and the true parallax could be significantly larger. Unfortunately, Gaia does not provide a parallax for this binary. Our orbital elements have previously been announced in the Information IAUDS No. 179 (Rica 2013a).

\begin{figure}[t]
\centering
\includegraphics[width=1\linewidth]{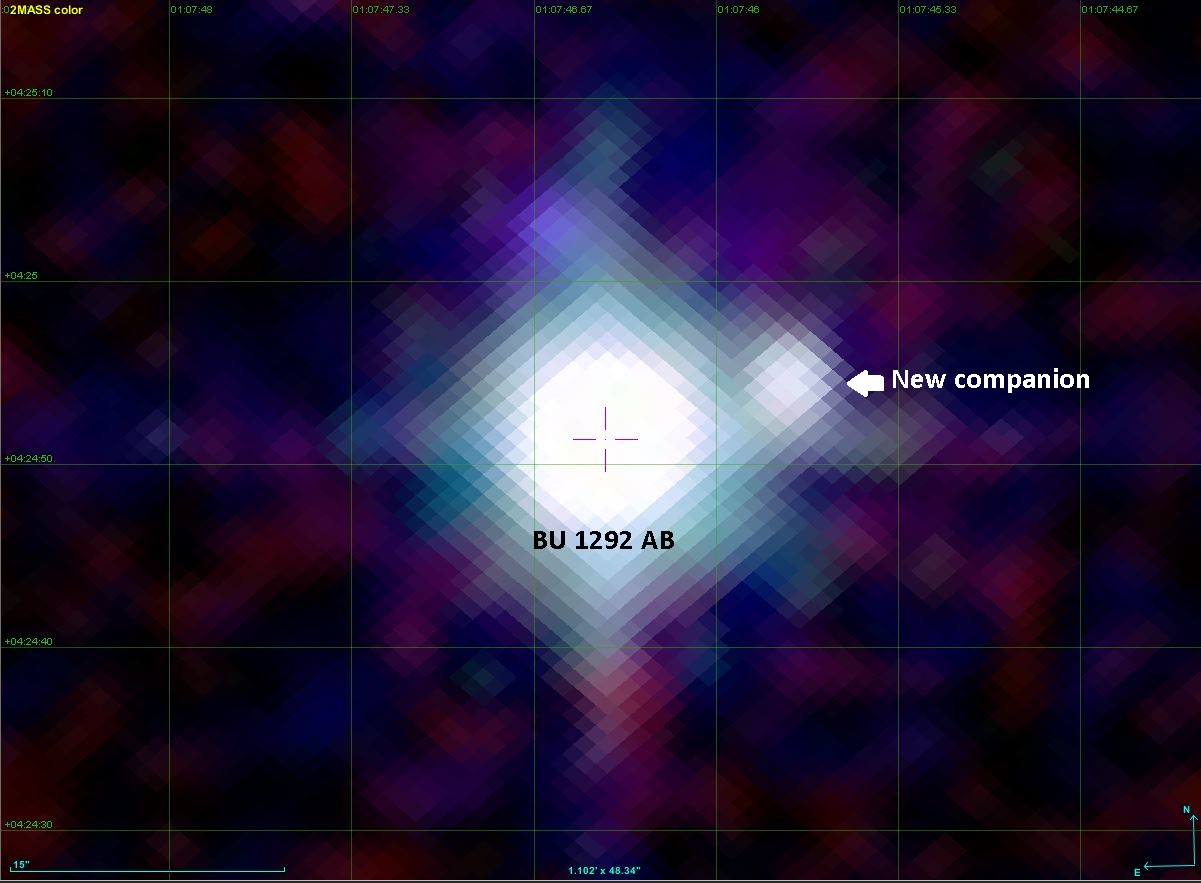}
\caption{New faint companion detected close to the binary star BU 1292. This image shows an RGB colour composition from J-, H- and K-band of 2MASS survey.}
\label{fig_1}
\end{figure}

In the literature, we find  a combined SpT of F5 (Cannon and Pickering 1918-1924) and F5/6V (Houk \& Swift 1999). LAMOST-DR5 catalogue (Luo et al. 2019) lists \emph{T}$_{eff}$ of about 6450 K, log \emph{g} = 4.1, and a metallicity [Fe/H] of $-0.08 \pm 0.01$. Using our tool ``Binary Deblending v5.0'', the component A seems to start to evolve. The fundamental parameters determined by us (individual SpT of F6IV-V + F5V, \emph{T}$_{eff}$ of approximately 6400 and 6500 K) are in excellent agreement with the literature. Our estimated age of $3.2_{-0.8}^{+1.1}$ Gyr is the first one published in the literature. Table 5 shows the good agreement between the unreddening observed and synthetic photometry (their difference is listed in the last column). In this study we present the first galactocentric velocity for this system, (\emph{U}, \emph{V,} \emph{W}) = (+31, -19, -13) km s$^{-1}$, which is consistent with a galactic thin disk population of young-medium age ($3-4$ Gyr), which agrees with our isochronal age estimation. The dynamical parallax was utilized for our astrophysical study because of the possible low quality value reported by Hipparcos. The dynamical stellar masses  agree with our findings (1.27 and 1.16 $M_\odot$). From the 3$^{rd}$ Kepler Law, using our orbital parameters, and the Hipparcos parallax, a dynamical total mass of about 12 $M_\odot$ is obtained. This is an unrealistic value which reinforces our suspicion about the Hipparcos parallax.

At a separation of 10.48” in direction of 285.7 deg (Gaia DR3), a faint star with a \emph{G} = 17.05 mag (\emph{V} = 17.8-17.9 mag) is found to share the same proper motion as the AB binary system. Figure \ref{fig_1} presents a 2MASS color image with the new weak companion. Gaia DR3 provides a parallax of $4.51 \pm 0.10$ mas, marginally comparable to our dynamical parallax for BU 1292 ($5.2_{-0.1}^{+0.4}$ mas), if we take into account that the formal errors listed in the Gaia catalogs are underestimated (El-Badry et al. (2021), Ren et al. (2021)). Gaia lists a proper motion for this companion of $\mu_\alpha cos \delta$ = $-18.52 \pm 0.13$ and $\mu_\delta$  = $37.83 \pm 0.11$ mas yr$^{-1}$. 

Using the relation deﬁned in the ``Modern Mean Dwarf Stellar Colour and Eﬀective Temperature Sequence'' (Pecaut \& Mamajek 2013) , we calculate a \emph{M}v = $11.08 \pm 0.13$, corresponding to a M3V star. Alternatively, from \emph{G} - \emph{J} colour vs SpT relation by Cifuentes et al. (2020) (see Figure 13 therein), we estimate a SpT of M3-4V for the new companion. \\

\begin{table}[t]
\caption{Comparison of combined observed and synthetic photometry for BU 1292.}
\label{Table_5}
\begin{tabular}{p{0.1\textwidth} p{0.17\textwidth}  p{0.2\textwidth} p{0.17\textwidth} r}
\hline
      & Observed photometry & Source  & Synthetic photometry & Diff. \\ 
\hline      
B     & 9.80 ± 0.04         & HIP     & 9.78            & 0.02  \\
V     & 9.31 ± 0.02         & HIP     & 9.31            & 0.00  \\
I     & 8.75 ± 0.03         &         & 8.75            & 0.00  \\
J     & 8.39 ± 0.03         & 2MASS   & 8.38            & 0.01  \\
H     & 8.17 ± 0.03         & 2MASS   & 8.18            & -0.01 \\
K     & 8.09 ± 0.03         & 2MASS   & 8.13            & -0.04 \\
B - V & 0.49 ± 0.04         & APASS   & 0.47            & 0.02  \\
V – I & 0.56 ± 0.04         & APASS   & 0.56            & 0.00  \\
V - K & 1.22 ± 0.04         &         & 1.18            & 0.04  \\
J - H & 0.22 ± 0.04         &         & 0.19            & 0.03  \\
H - K & 0.08 ± 0.04         &         & 0.06            & 0.02  \\
J - K & 0.22 ± 0.04         &         & 0.19            & 0.03  \\
\hline
\end{tabular}
\end{table}

\underline{WDS 01417-1119 = STF 147 = ADS 1339} \\

STF 147 ( = HD 10453 = HIP 7916) was discovered by Struve (1837) in 1822 and consists of two bright stars with \emph{V} magnitudes of 6.1 and 7.3, separated by 0.1”. Since its discovery, this binary system has been observed 175 times covering approximately 254 degrees. Using this full dataset, we present the first orbital solution for STF 147. The periastron occurred in 2014 and the calculated orbit show a high eccentricity and inclination. The orbital elements have previously been announced in the IAUDS No. 181 (Rica 2013c).

STF 147 presents IR, UV and X emissions. Literature reports a large amount of references with combined SpT for this system ranging from F4V to F6V. Some of these references are Adams et al. (1935) who reported F2V+F3V SpT for the components, and the Bright Star Catalog, which lists SpT of F5+F7. More recently, Gray et al. (2006) determined a combined SpT of F6V. Related with metallicity, Casagrande et al. (2011) reported a metallicity [Fe/H] of -0.38 while Marsakov et al. (1995) determined a value of -0.26. 

Using \emph{U} – \emph{B} color (Wild 1969), the Hipparcos \emph{BVI} photometry, the 2MASS \emph{K}-band photometry (\emph{J} and \emph{H} magnitudes are  saturated)  as  well  as a magnitude  difference  $\Delta$\emph{V} = 1.13 ± 0.12 mag (average values obtained from the WDS database for wavelengths of about 550 nm), and applying our tool ``Binary Deblending v5.0'', we find an optimal solution with a minimum $\chi^2$ represented by stars showing SpT of F4IV-V and F7V. We also obtain a metallicty of [Fe/H] = -0.26, which are in agreement with the literature. We estimate an age of $3.5_{-0.8}^{+1.1}$ Gyr for this system. The dynamical parallax and masses are consistent with the Hipparcos parallax and the expected total mass. 

In January 1991, the ROSAT satellite detected an X-ray source (2RXS J014144.9-111940) located 14” south of STF 147. A faint star (Gaia DR3 2463529782844691712) with a \emph{G} = 15.64 mag, is located at 38” toward southwest of the X-ray source. The positional error listed in the ROSAT catalogue is 11”. The log fx/fv of the A and B stellar components marginally match with that of F-type stellar objects (Table 1 in Agüeros et al. 2009). Krautter et al. (1999) determined a 90\% error circle of approximately 30” for the distance between the optical counterpart and the ROSAT X-ray astrometric position. Greiner \& Richter (2015) detected 256 probable optical identifications of the 370 ROSAT sources within the astrometric error circle of ROSAT. For all exposed here, HD 10453 is the most likely optical counterpart in agreement with Flesch (2016) who gives a 65\% probability, and Salvato et al. (2018) who estimated a probability of 99.7\% to be the actual optical counterpart. The X-ray luminosity, Log (Lx) = 28.63, is typical of Hyades stars, leading to the conclusion that STF 147 has an age similar to the Hyades cluster, approximately 0.6 Gyr. However, this age contradicts the kinematic age and the age calculated using theoretical isochrones in this work. The galactocentric velocity (Casagrande et al. 2011) suggests that this system belongs an old disc stellar population. The discrepancy between the age, based on X-ray emission, and the age based on kinematic or other methods, could be due to an unresolved binary in STF 147 or by an evolved star like the A component (Suchkov et al. 2003).  \\

\underline{WDS 04573+5345 = D 5 AB = ADS 3536} \\

WDS 04573+5345 (7 Cam = HD 31278 = HIP 23040) is a quadruple system located at a distance of 106 pc (Gaia DR3). It is composed of an SB1 binary (components Aa, Ab) with an orbital period of 3.884 days. The B component is at approximately 0.6” and it is listed as D 5 AB stellar system in the WDS catalogue, composed of stars with \emph{V} magnitudes of 4.6 and 7.8. The wide C component, with \emph{V} = 11.2 mag, is separated at approximately 26”. According to the Gaia DR3 catalogue, it shares a common parallax and proper motion with D 5 AB, suggesting a physical binding. The component A is catalogued as the suspected variable NSV 16198, a rotating ellipsoidal variable. In this study new orbital parameters are presented for AB components. Our orbital solution was previously announced in the Information IAUDS No. 179 (Rica 2013a).

The B component was first detected and measured by Dembowski in 1865 (Dembowski 1884). Since then, a total of 31 measurements have been reported, covering an arc of 107 degrees. Baize (1979) calculated orbital parameters and it is worth noting that measurements before 1983 considered the faint component as the A member. In this work, we reverse (adding 180 deg) the quadrant for these measures. The recent measurements show a closely linear trajectory that deviates significantly from Baize‘s orbit. 

We conduct a thorough dynamical study to determine the gravitational binding between the A and B components. The relative and projected velocity (\emph{V}$_{tan} = 5.40 \pm 0.38$ km s$^{-1}$) is nearly half the maximum escape velocity (\emph{V}$_{esc-max} = 12.31$ $km s^{-1}$), indicating that D 5 AB is gravitationally bound. Our Monte Carlo approach confirmed this result, with 100\% of simulations showing \emph{V}$_{tan}$ < \emph{V}$_{esc-max}$. 

We find in the literature that 7 Cam presents a SpT ranging from B9.5V to A1V (Levato \& Abt 1978; Gray \& Garrison 1987; Abt 1995)). Pickles \& Depagne (2010) listed a IV luminosity class, A0IV for this star, likely influenced by the multiple nature of 7 Cam. The Gaia DR3 catalogue lists D 5 AB as an unresolved object and its brightness causes a low-quality astrometric solution due to saturation and other different facts. We study this system using our tool ``Binary Deblending v5.0'' with PARSEC isochrones of solar metallicity, as no [Fe/H] values were available in the literature. Our analysis yielded a minimum $\chi^2$ solution that matches the 0.2 Gyr isochrones. Based on our study, the component A, the SB1 binary, consist of B9IV-V + F5/6V stars, while the components B and C are classified as F0V and K0V stars, respectively. 

Marton et al. (2016) listed the C component in a list of likely Young Stars Objects but no ages were provided. King et al. (2003) studied the potential membership to Ursa Majoris group and concluded that it is likely a non-member. Riedel et al. (2017) also included this object in their Catalog of Suspected Nearby Young Stars. In this study, by utilising the BANYAN $\Sigma$ online tool , we determine a 0\% probability for C component to be a member of any of the young moving groups considered in BANYAN $\Sigma$. 

 The literature does not report radial velocities for AB or C. However, independently of this data, this system could have a kinematic behavior typical of young star populations.
 
About the spectroscopic orbit for the component A, Lucy \& Sweeney (1971) calculated an SB1 orbit with an orbital period of 3.88 days describing an almost circular motion. Using the expressions 3 and 5 in Abushattal, Docobo \& Campo (2020) and the process described by Docobo \& Andrade (2006), we determine a tentative $\Delta$\emph{V} = 4.5 mag, a semimajor axis of $0.081 \pm 0.006$ UA ($0.76 \pm 0.05$ mas) and an orbital inclination of $34.6 \pm 3.9$ deg. 

The total mass of the system, discarding the C component, is estimated to be about 6.6 $M_\odot$. However, the total mass obtained from the initial orbital solution published in IAUC, and the parallaxes of Hipparcos and Gaia DR3, provide values of 4.5 and 3.6 $M_\odot$, which is significantly smaller than expected based on our analyses. As a result, we have recomputed and improved our initial orbital solution, leading to a total mass of about 6.6 $M_\odot$. Our new orbital solution, has a much shorter orbital period (824 years) compared to the initial solution (2700 years). However, it should be noted that this is a preliminary solution, with only the inclination well-constrained, i = $106.2_{-7.8}^{+7.2}$ deg. \\

\underline{WDS 08031-0625 = A 1581 = ADS 6547 } \\

A 1581 (= HIP 39383 = BD-06 2423) is a triple system composed of a close astrometric binary (the Aa, Ab components), discovered recently by Gaia satellite, and the wide component B, separated by 1.4”.  A 1581 is a system with moderated proper motion, –56.2 mas yr$^{-1}$ in RA and +45.5 mas yr$^{-1}$ in DEC, derived  from Tycho-2 and located at a distance of 68.7 pc (Gaia  DR3).

Regarding AB componnets, Baize (1961) published an orbit with a period of 94.7 years while Heintz (1967b) calculated another orbit with a period of 267 years. Since then, large residuals have been observed. The orbit presented in this work has a significant longer orbital period of 1225 years respect that obtained by Heintz. Izmailov (2019) reported a new orbit which yields a dynamical total mass of 15-16 $M_\odot$, very much higher than the expected one.

A 1581 AB shows a very linear relative motion. Our dynamical study demonstrates that the relative total velocity is smaller than the escape velocity, indicating that A 1581 AB is a likely gravitationally bound system. The Gaia DR3 catalogue resolved the A and B components at 1.440” providing common proper motions and parallaxes. 

Gaia re-analysed the astrometry of the primary component of A 1581. Using a non-single star orbital model for sources compatible with a two-body solution, they find the orbital parameters reported in Table 6. 

The photocentre and barycentre lie along the line connecting the Aa and Ab components. The ratio of the distance between the barycentre and Aa to the distance ``Ab-Aa'', and the ratio of the distance between the phtotocentre and Aa to the distance ``Ab-Aa'', are defined by the well-known parameters B and $\beta$. 

The ratio of the relative semi-major axis (that is, expressed as the Ab component with respect to Aa) and photocentric semi-major axis (that is, the photocentre with respect to the barycentre) depends on the mass ratio and the magnitude difference of the components (read Gontcharov \& Kiyaeva 2002, section 2). 

We apply an iterative process to determine a $\Delta$\emph{V} = 0.63 mag, initial masses of 0.78 and 0.71 $M_\odot$ for Aa and Ab components, respectively, and a relative semi-major axis (a) of 4.1 mas (0.28 UA).

This stellar system has a combined SpT of K0V (Bastian \& Roeser 1993) while we obtained here, from \emph{BVIJHK} magnitudes, a G8V type. We study this system using PARSEC isochrones with metallicity, [Fe/H], in the range from -0.1 to -0.4. We obtain a minimum $\chi^2$ solution with a metallicity [Fe/H] of -0.38 and with SpT of G9V and K2V for Aa and Ab while the B component is classified as a G8V star. The LAMOST DR5 catalogue (Luo et al. 2019) lists a metallicity of $0.00 \pm 0.02$, probably due to the multiple nature of HIP 39383, while Gaia DR3 reports a metallicity of $-1.08 \pm 0.04$ for the primary component surely in error.

The use of PARSEC isochrones yield a total mass of $2.2 \pm 0.1$ $M_\odot$. Our orbital solution has a total dynamical mass (about 1.7 $M_\odot$) lower than that calculated in our astrophysical study. Thus, an improvement of this parameter is clearly needed.

In this work, we present the first galactocentric velocity for this system, (\emph{U},\emph{V},\emph{W}) = (-39.9, -11.6, -1.5) km s$^{-1}$, which matches with a galactic thin disk population of young-medium age (3-4 Gyr). Using the BANYAN $\Sigma$ on-line tool  we find a 0\% probability for the C component to be a member of any of the young moving groups considered in BANYAN  $\Sigma$. \\

\begin{table}[t]
\caption{Gaia photocentric orbit for the A component of A 1581.}
\label{Table_6}
\begin{tabular}{lll}
\hline
Parameter & Value  & Error \\
\hline
P [yr] & 0.1195 & 0.0002 \\
P [days] &  43.635 & 0.084 \\
T {[}yr{]}   & 2016.038 & 0.014       \\
e            & 0.575 & 0.298          \\
$\alpha$ {[}mas{]}  & 0.473 & 0.016          \\
i {[}deg{]}     & 69.3 & 6.0          \\
$\omega$ {[}deg{]}    & 293.8 & 16.9  \\
$\Omega_{2000}$ {[}deg{]}   & 51.1 & 4.9 \\
\hline
\end{tabular}
\end{table}

\underline{WDS 08231+2001 = HO 525 AB = ADS 6776 } \\

HO 525 AB ( = HD 70492 = HIP 41098) was discovered by Hough (1899) in 1895 and consists of a pair of nearly twin stars with \emph{V} magnitudes of 9.7 and 9.8, separated by 0.4” and located at a distance of 141 pc. In total, it has been measured on 38 times. 

In our orbital calculation, we consider our 5 astrometric points published in the Paper I. For the measures of 2007.8186 and 2008.15, we appropriately reverse the quadrant of $\theta$. The astrometric measures cover an arc of approximately 198 degrees. In comparison to the previous orbit presented by Baize (1994), our results show an important reduction in the residuals for high-resolution measures, leading to a revised orbital period of 935 years. Previous orbital calculations provide 121.8 years. Our orbital solution was previously announced in the IAUC No. 181 (Rica 2013c).

The WDS catalogue lists a wide stellar component with a  \emph{V} = 13.3 mag, located at about 38” from AB. Based on Gaia DR3 data, we conclude that this wide companion has no physical relation with the AB system.

In the literature, combined SpT of F6V (Pickles \& Depagne 2010) and F6IV-V (Stephenson \& Sanwal 1969) have been reported for HO 525 AB. Gaia DR3 lists HIP 41098 with a complete astrometric solution with RUWE of 7.6, which might be affected by duplicity. 

Past analyses, such as the APOGEE-2 DR16 catalogue (J\"ohnsson et al. 2020), suggested a metallicity, [Fe/H] = $0.00 \pm 0.02$, and Sprague et al. (2022) reported a value of $-0.10 \pm 0.01$. On the other hand, the TESS Input Catalog version 8.2 (Paegert et al. 2021) listed a metallicity [M/H] of $-0.04 \pm 0.01$. However, these works do not consider HIP 41098 as a binary star. This fact might impact in the accuracy of their metallicities estimation.

Our analysis found the best fit with respect to the observed combined photometry in \emph{BVIJHK} bands for isochrones with a centroid metallicity [Fe/H] = -0.26 and an age of 0.4 Gyr. We constrained the system age to be younger than 3.6 Gyr at a 95\% confidence interval (at 2$\sigma$ level). We conclude that this stellar system is composed of twin stars showing both a F4V spectral class. The total dynamical mass (1.9 $M_\odot$) is in agreement with the mass obtained from our isochronal fitting (1.1 + 1.1 $M_\odot$). The dynamical parallax is in agreement with the Hipparcos and Gaia DR3 parallax. Additionally, we present the first galactocentric velocity for this system,  (\emph{U},\emph{V},\emph{W}) = (-5.8, -8.9, -4.0) km $s^{-1}$, which corresponds to a galactic thin disk population of young-medium age (3-4 Gyr). \\

\underline{WDS 08286+3502 = WOR 19 } \\

WOR 19 (GJ 308 = HIP 41554) is a binary system composed of faint and red dwarf stars (\emph{V} magnitudes of 11.45 and 11.58) located at 20 pc. These stars show a high proper motion of 1.05 arcsec yr$^{-1}$ and are separated by 0.5”.

The quadrants for measures performed between 1978 and 1988 and the measure of Hipparcos were reversed for our orbital calulation. Couteau (1982) computed the last known orbit, and our orbital solution, which was previously announced in IAUDS No. 178 (Rica 2012b), presents a longer orbital period (50.14 years) in comparison to the earlier orbit (32.70 years). The astrometric data cover an entire revolution. While the dynamical parallax (45.5 mas) agrees with the Hipparcos one ($50.8 \pm 6.3$ mas) within errors margins, the dynamical mass falls very close to the expected values. Gaia DR3 classifies this object as a long-period variable star and lacks an astrometric solution.

We apply our tool ``Binary Deblending v5.0'' yielding individual SpT of M1.5V and M1.5V \footnote[3]{For this binary star, we utilized the \emph{U}, \emph{B}, and \emph{V} photomertry of Mermilliod (1991) in addition to the 2MASS photometry}. 

The astronomical literature reports combined SpT ranging from M0V (Buscombe 1998) to M1.9V (Pickles \& Depagne 2010). Our \emph{T}$_{eff}$ = $3665 \pm 30$ K and $3648 \pm 50$ K, for the A and B components, agree with that obtained by Gaidos (2014), $3700 \pm 24$ K, and Muirhead et al. (2018), $3683 \pm 92$ K. However, our result for the metallicity, [Fe/H] = $-0.40 \pm 0.13$, falls within a 95\% confidence interval of -0.60 to -0.24, which is significantly lower than Gaidos's value, $+0.45 \pm 0.12$. This discrepancy could be attributed to the challenging photometry of HIP 41554. 

In this work, we also present the first estimation of the galactocentric velocity for this system, and we obtain (\emph{U},\emph{V},\emph{W}) = (-79.0, -21.1, -62.3) km s$^{-1}$ that matches with the typical velocity of the galactic thin/thick disk population. \\

\underline{WDS 12154+4008 = A 1999 = ADS 8480 } \\

A 1999 ( = HD 106592 = HIP 59776 = NLTT 30139) is a high proper motion stellar system with \emph{V} = 9.0 and 11.0 mag, located at a distance of 40 pc (Gaia DR3) and separated by approximately 0.17”. This binary system was discovered by Aitken (1909) and has been observed 33 times. The measurements cover an orbital arc of approximately half a revolution. The orbital path of this system is closing significantly and displaying important discrepancies when it is compared with the previous orbit calculation by Seymour et al. (2002). Our newly derived orbit, previously announced in the IAUDS No. 180 (Rica 2013b), reveals an orbital period (200 yr) shorter than the previous orbital solution (630 yr), with a periastron occurring in 2014.1.



\begin{figure*}[th!]
\centering
\begin{tabular}{cc}
    \includegraphics[width=0.50\linewidth]{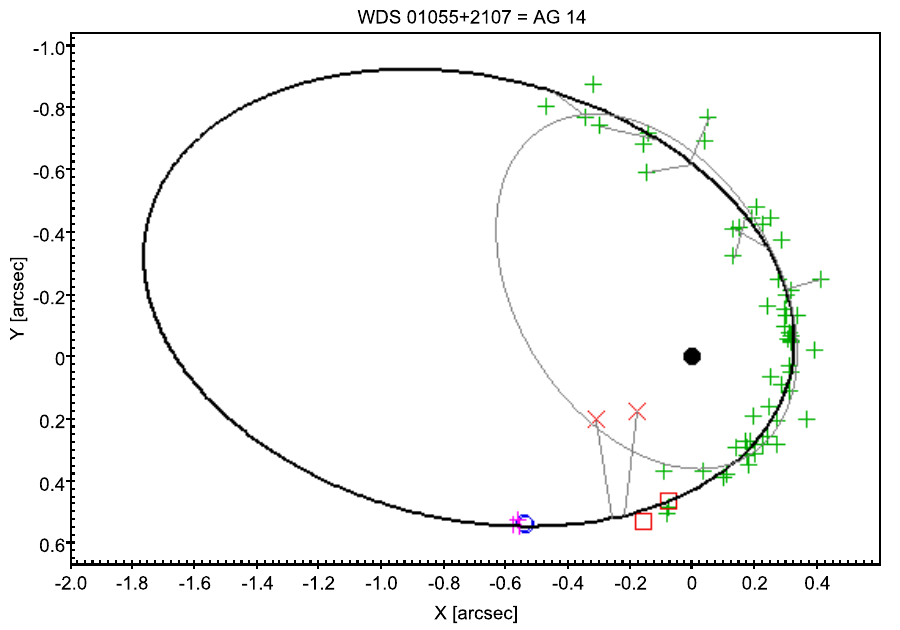}&
    \includegraphics[width=0.47\linewidth]{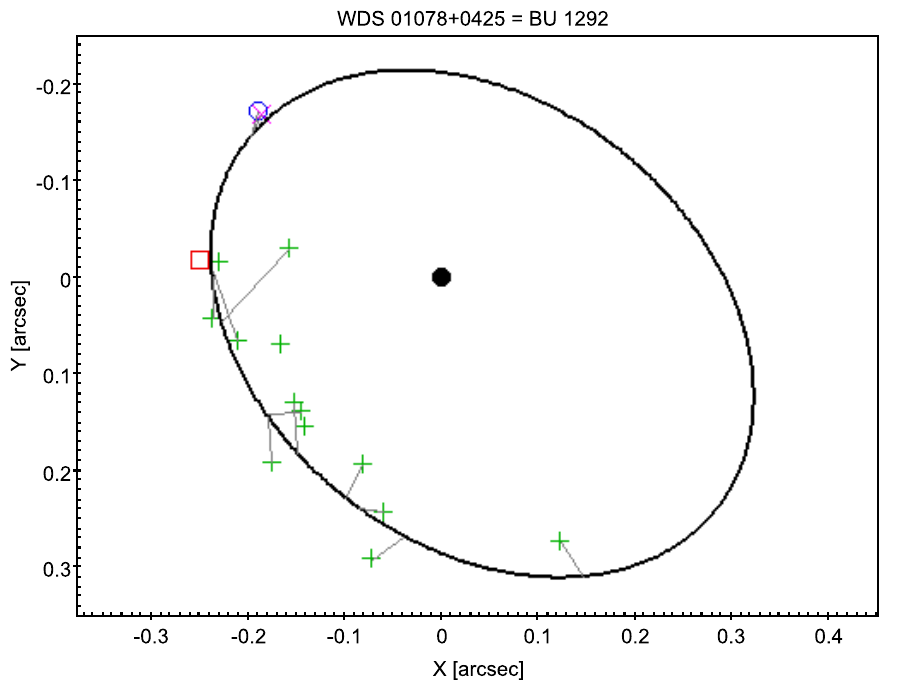}\\[2\tabcolsep]
    \includegraphics[width=0.47\linewidth]{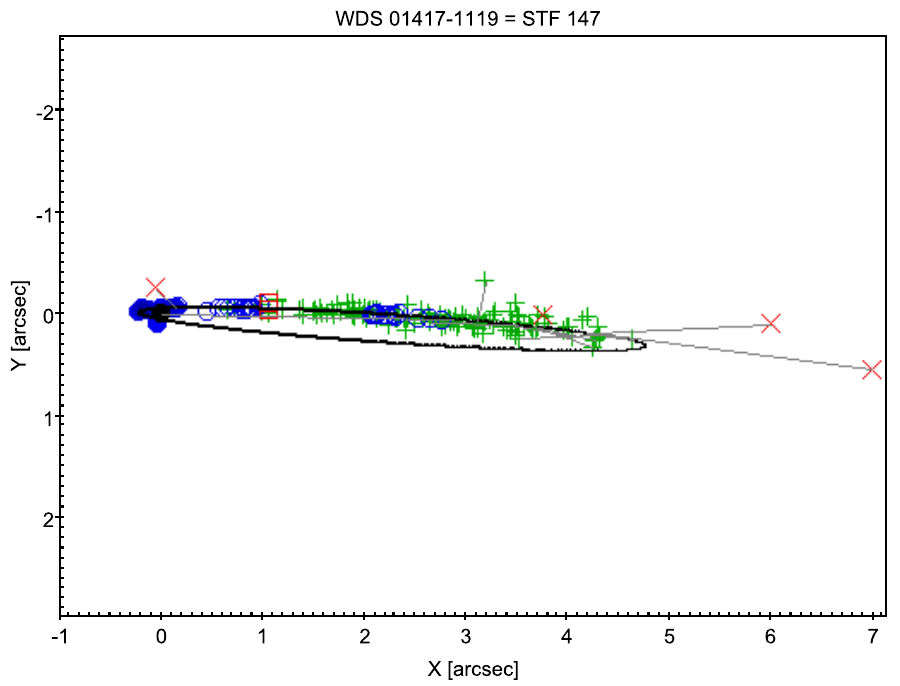}&
    \includegraphics[width=0.47\linewidth]{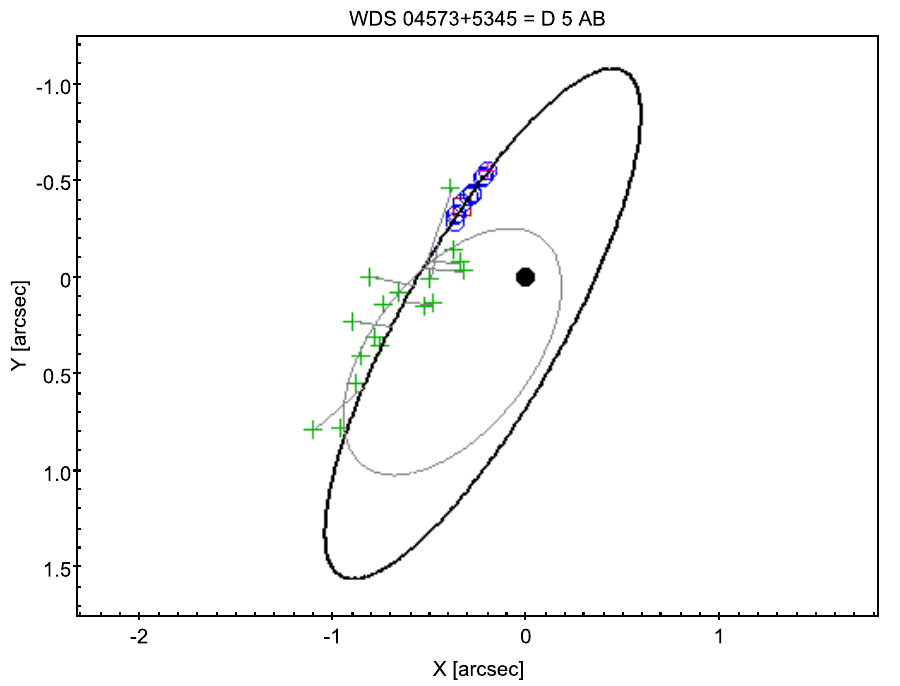}\\[2\tabcolsep]
    \includegraphics[width=0.47\linewidth]{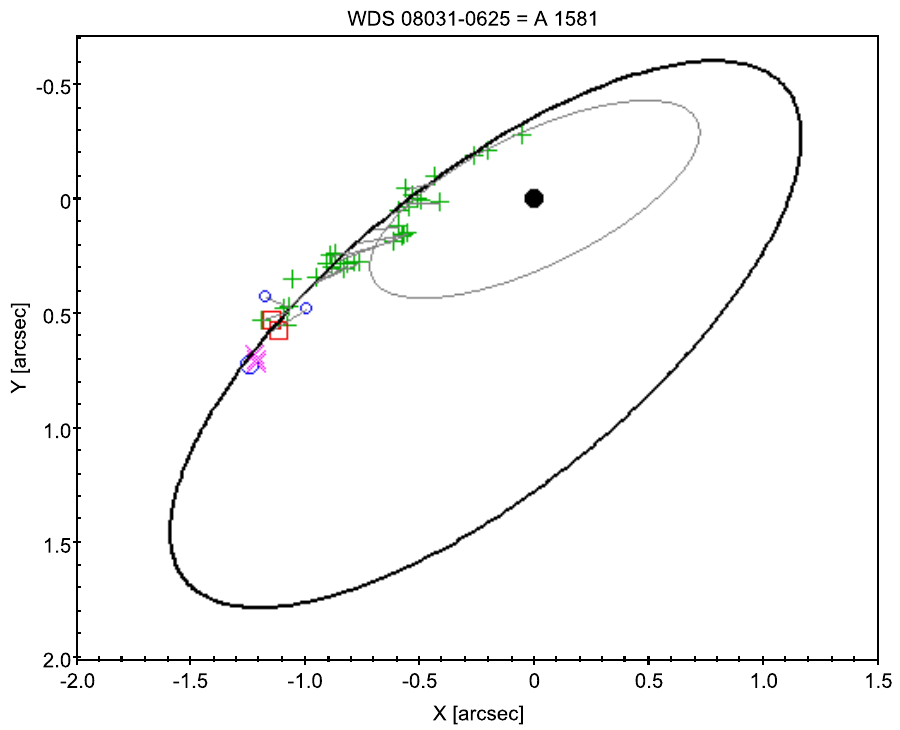}&
    \includegraphics[width=0.47\linewidth]{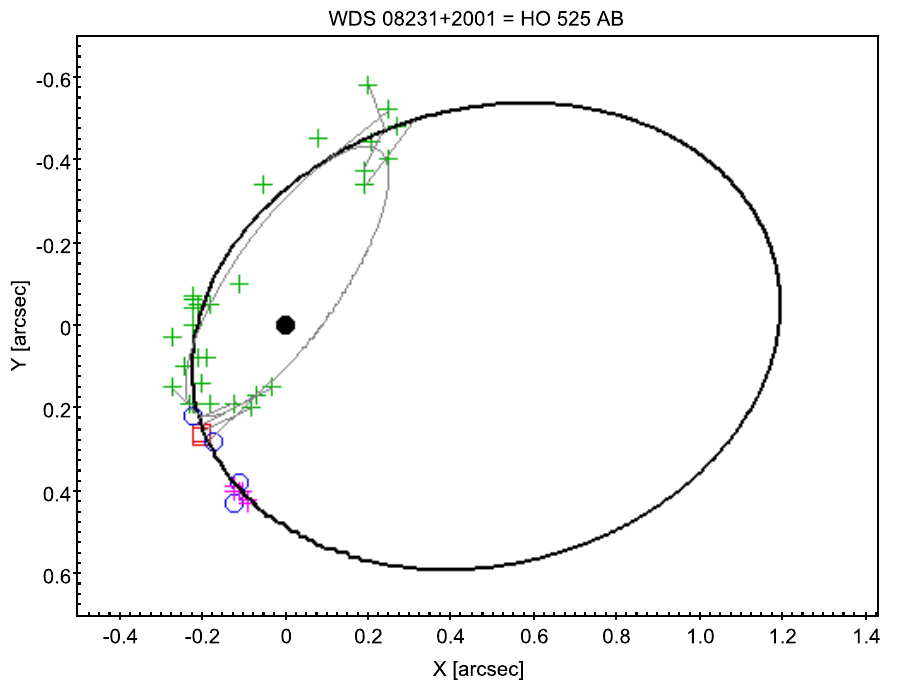}    
\end{tabular}
\caption{Orbits of AG 14, BU 1282, STF 147, D 5 AB, A 1581,and HO 525.}
\label{fig_2}
\end{figure*}

\begin{figure*}[th!]
\centering
\begin{tabular}{cc}
    \includegraphics[width=0.47\linewidth]{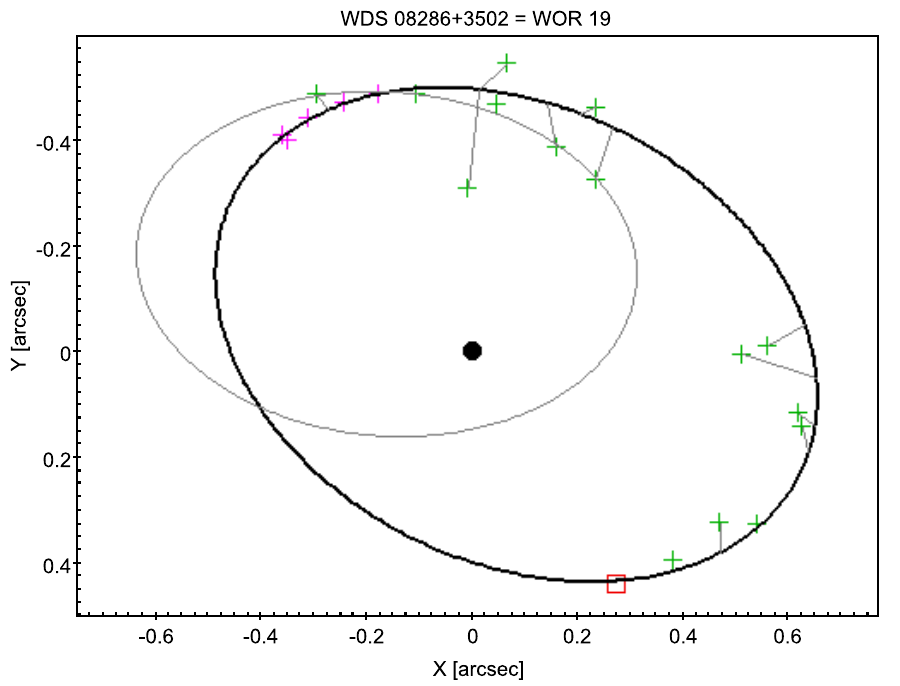}&
    \includegraphics[width=0.45\linewidth]{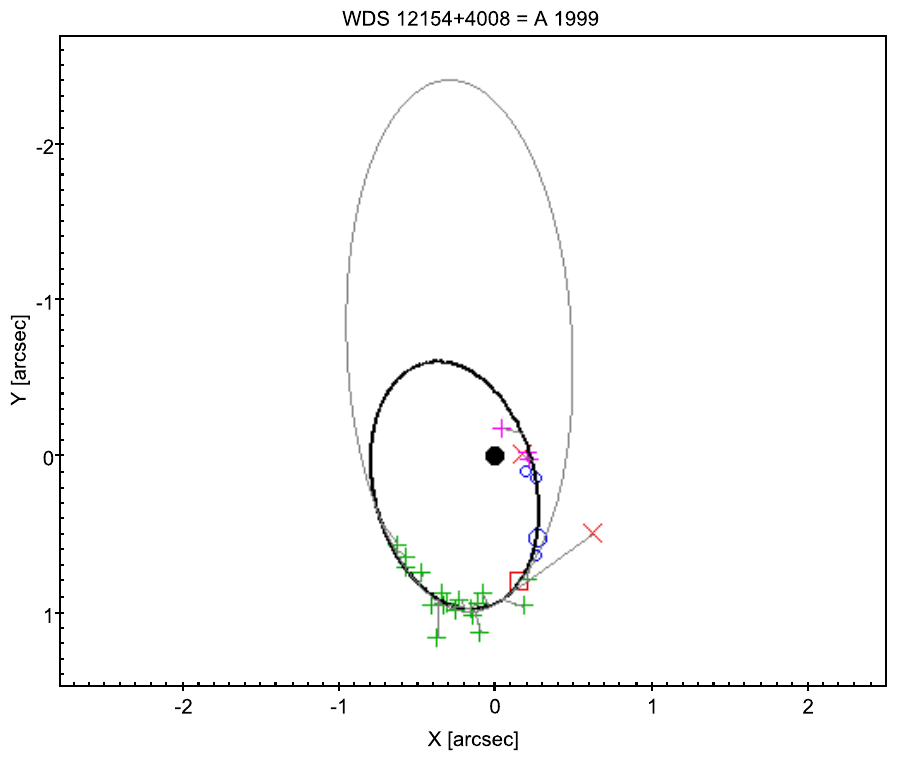}\\[2\tabcolsep]
    \includegraphics[width=0.47\linewidth]{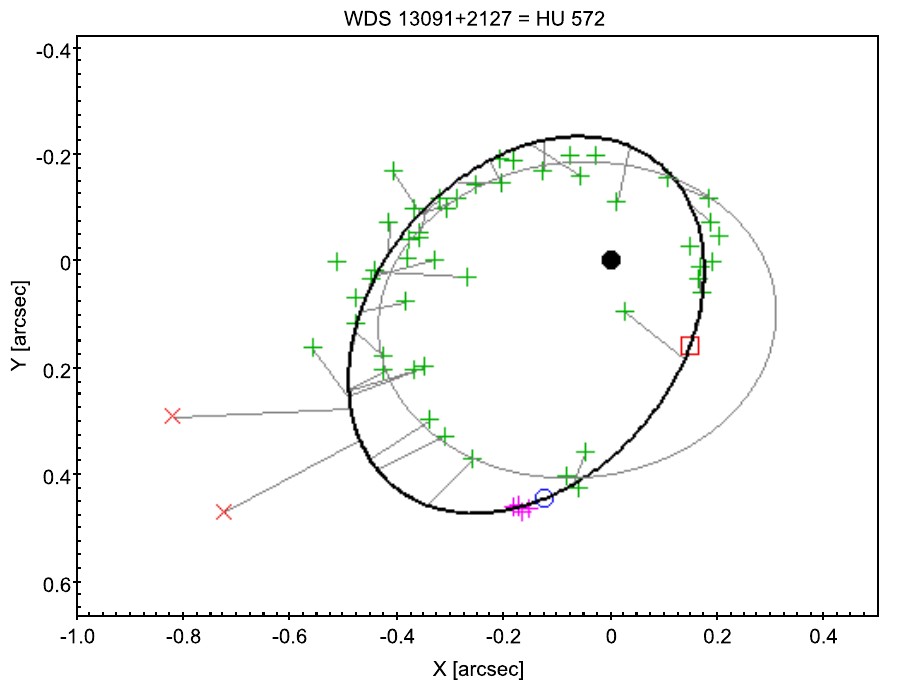}&
    \includegraphics[width=0.47\linewidth]{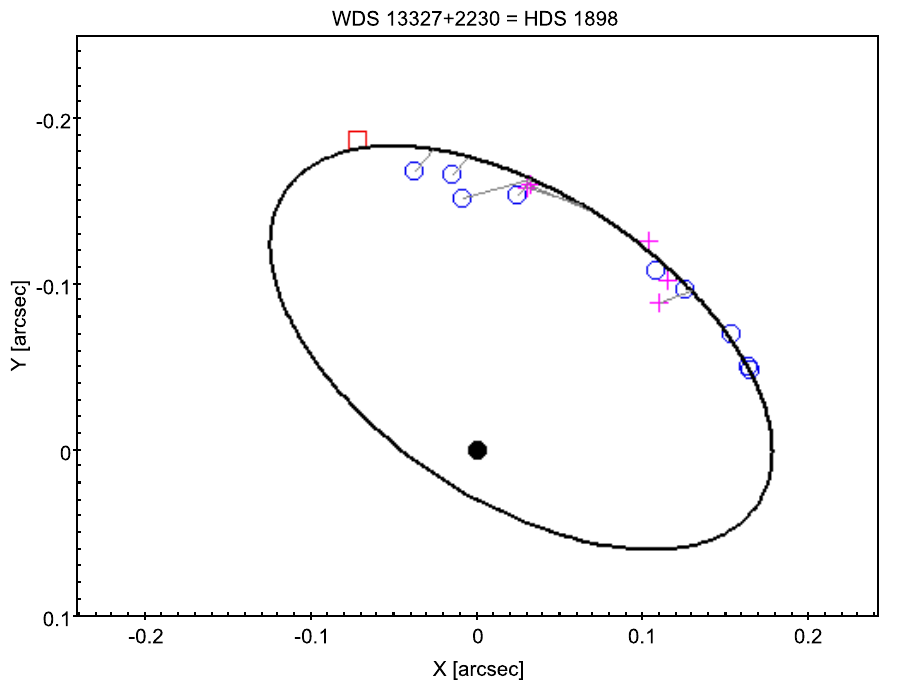}\\[2\tabcolsep]
    \includegraphics[width=0.47\linewidth]{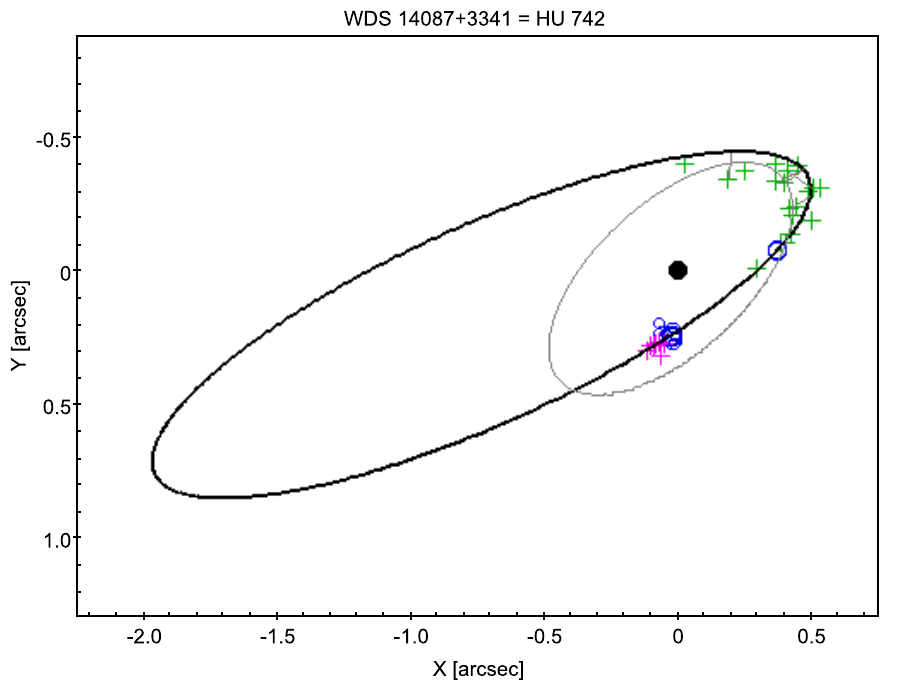}&
    \includegraphics[width=0.47\linewidth]{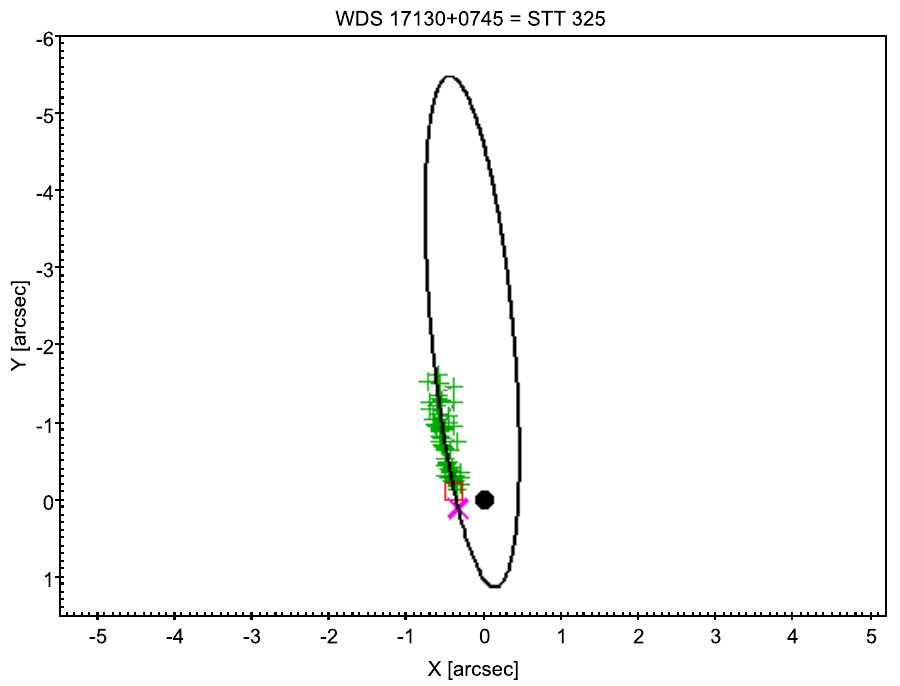}    
\end{tabular}
\caption{Orbits of WOR 19, A 1999, HU 572, HDS 1898, HU 742, and STT 325.}
\label{fig_3}
\end{figure*}

\begin{figure}[th!]
\centering
\begin{tabular}{c}
    \includegraphics[width=1\linewidth]{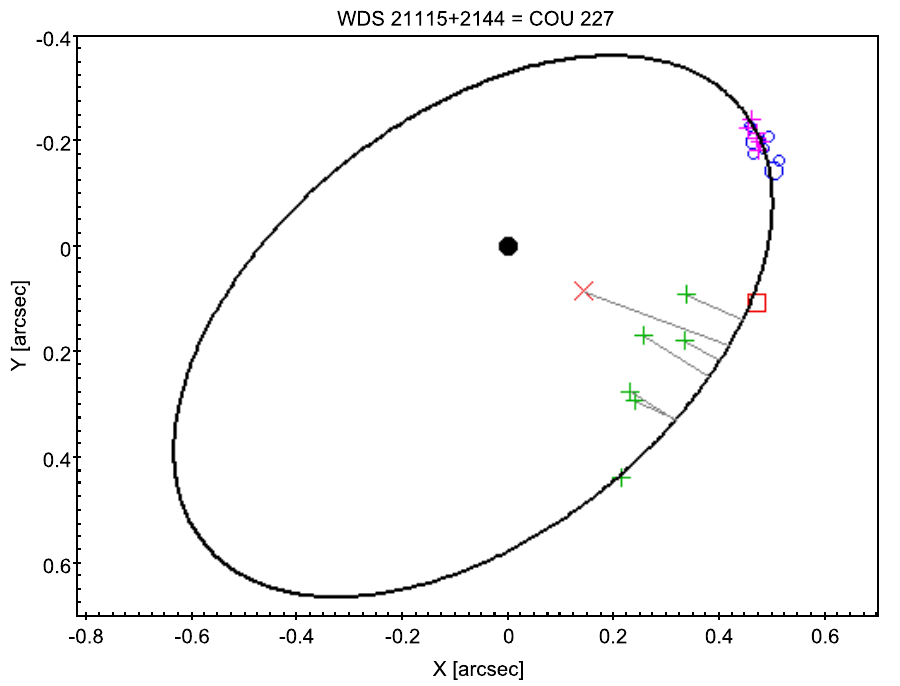}\\[2\tabcolsep]
    \includegraphics[width=1\linewidth]{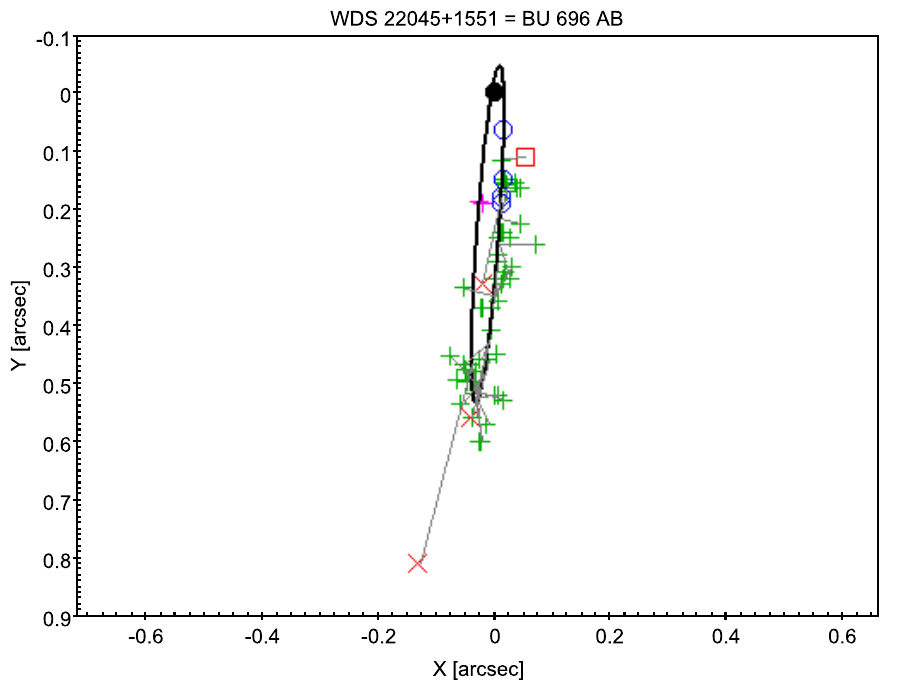}\\[2\tabcolsep]
    \includegraphics[width=1\linewidth]{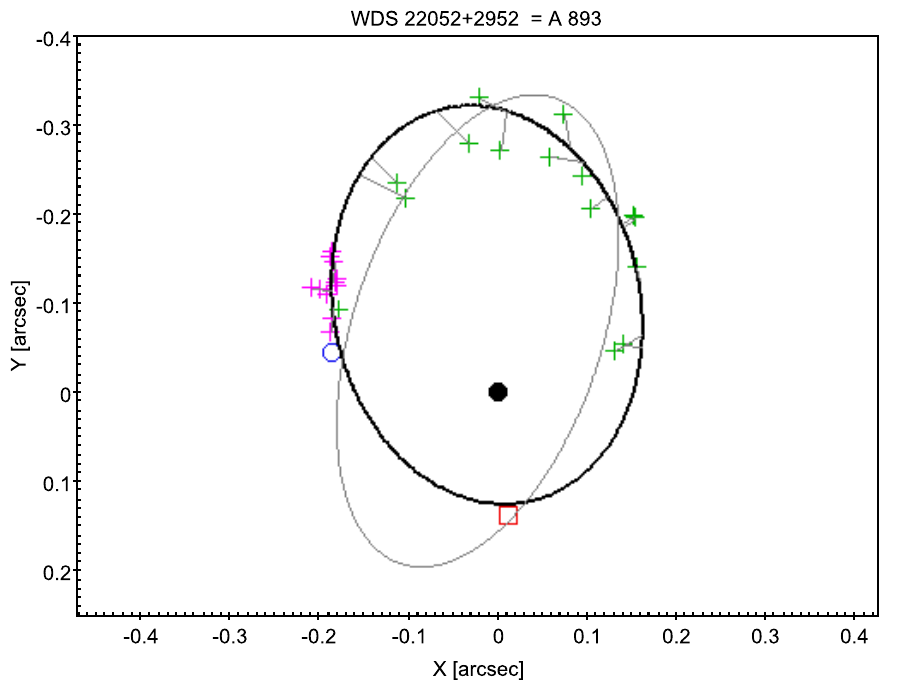}
\end{tabular}
\caption{Orbits of COU 227, BU 696 AB, and A 893.}
\label{fig_4}
\end{figure}

Upgren (1962) incorrectly reported a combined SpT of G9IV (data listed in the WDS catalogue and in the Simbad database) whereas Skiff (2007) reported a SpT of K2V. Our comprehensive astrophysical analysis concludes that this stellar system presents a metallicity [Fe/H] > -0.10 (at 95\% confidence level). It is composed of K1.5V and K6V stars, with stellar masses in agreement with the dynamical masses.

We use the Hipparcos parallax for our investigation of this system. Our \emph{T}$_{eff}$, 5151 and 4286 K for the A and B components, are consistent with those presented in Gaia DR3 Part 1, approximately.

Gaia DR3 provides an astrometric solution for this target, but with a high RUWE value of 16.98. This is likely influenced by the duplicity of A 1999, as the physical separation between its components was 33 AU in 1991.25 and 11 AU in 2016.0. Kervella, Arenou \& Th\'evenin (2022), using Hipparcos and Gaia data, detected a significative proper motion anomaly. Taking into account the physical separation of the components, the binarity of A 1999 could be the cause of this anomaly.

In this work, we use the Tycho-2 proper motion and the radial velocity from Gaia DR3 ($+32.78 \pm 1.03$ km s$^{-1}$) to estimate, for the first time, the galactocentric velocities, (\emph{U},\emph{V},\emph{W}) = (-36, -19, +28) km s$^{-1}$. These velocities correspond to a population within the galactic thin disk of young-medium age (3-4 Gyr). This object is not a member of any of the young moving groups considered in BANYAN $\Sigma$. \\

\underline{WDS 13091+2127 = HU 572 = ADS 8799} \\

This pair, HD 114255 = HIP 64175, is composed of stars with magnitudes \emph{V} = 8.7 and 9.7, located at a distance of 114 pc (Gaia DR3). It was discovered by Hussey (1902) and has been astrometrically measured 60 times, covering more than one revolution. The previously published orbit by Baize (1985) displayed significant discrepancies in comparison to recent measurements. Our updated orbital elements have previously been announced in the IAUDS No. 176 (Rica 2012a).

Cannon \& Pickering (1918-1924) listed a combined SpT of G5, whereas Yoss and Griffin (1997) reported G3V with a metallicity, [Fe/H] = +0.06. In addition to these values, regarding to the metallicity, other authors indicated values of -0.12 (Casagrande et al. 2011) and -0.08 (Ammons et al. 2006). The parallax measured by the Hipparcos satellite is $12.75 \pm 1.43$ mas while Gaia satellite (DR3 Realease) reported $8.79 \pm 0.30$ mas, which also indicates a high RUWE value, 9.495, suggesting an astrometric solution significantly affected for the binary nature of HD 114255. 

We do not use the Gaia DR3 parallax in our calculation because the dynamical mass significantly exceeds the expected value. Instead, we use the Hipparcos parallax to estimate our astrophysical data. 

We obtain a $\Delta$\emph{V} = $1.0 \pm 0.2$ mag using visual WDS historical observations (Hipparcos reported a $\Delta$\emph{Hp} = $1.3 \pm 0.2$). This value is an important input data for our astrophysical tool ``Binary Deblending v5.0''. When we use the Hipparcos parallax, our minimum $\chi^2$ solution yields a [Fe/H] = -0.04, approximately. We also obtain SpT of G0V and G8V (with masses of 1.09 and 0.94 $M_\odot$) for the primary and secondary components, respectively.  

Casagrande et al. (2011) determined a galactocentric velocity (\emph{U},\emph{V},\emph{W}) of (-4, -8, -1)  km s$^{-1}$, which corresponds to a star belonging to the young stellar population in the thin Galactic disk. This kinematic behavior was no identified with any known young kinematic group. \\

\underline{WDS 13327+2230 = HDS 1898} \\

HDS 1898 ( = HIP 66072 = BD+23 2581) was discovered by the Hipparcos satellite in 1991. It consists of a couple of stars with \emph{V} = 10.3 and 10.8 mag Subsequent measurements were not made until 16 years later, when Hartkopf observed it in 2007 (Hartkopf \& Mason 2009). In total, it has been measured on 16 times, with the most recent observation conducted in 2021, covering roughly one revolution. In this work, we reverse the quadrant of $\theta$ for the 2007 measurement and present the first orbital solution for this system. The calculated orbital period is 30.87 years with a periastron occurring in 1998. These orbital elements were previously reported in the IAUDS No 180 (Rica 2013b).

Buscombe et al. (1995) reported a combined SpT of K4Ve, whereas Tsantaki et al. (2022) estimated a combined \emph{T}$_{eff}$ of about 4500 K. However, Sprague et al. (2022)  and Xing \& Xing (2012) reported 4700 K.

The galactocentric velocity by Binks et al. (2020) is typical  of  stars  belonging  to  the  thin  disk population of young-medium age. Based on our astrophysical study, we conclude that this stellar system comprises K3V and K4.5V stars, with stellar masses agree with the dynamical mass. The ROSAT satellite observed HIP 66072 between July 1990 and January 1991, detecting significant X-ray emission. Additionally, Binks et al. (2015) reported an important equivalent width EW(\emph{Li}) of $142 \pm 14$ m\AA. Both values are indicators of the youth of this system.

 In our work, considering the binary nature of the system, we calculate an X-ray luminosity Lg (\emph{X}) = 29.9 erg s$^{-1}$. This value is higher than that of the Pleiades open cluster, suggesting an age slightly lower than 125 Myr. When examining the Figure 7 from Sestito \& Randich (2005), we find that HIP 66072's EW(\emph{Li}) falls slightly below the stars of open cluster with 30-50 Myr and slightly higher than the stars in the Pleyades with similar \emph{T}$_{eff}$. However, using similar figures from Guti\'errez et al. (2020), HIP 66072 has a EW(\emph{Li}) slightly lower than stars with similar \emph{T}$_{eff}$ in 30-40 Myr open clusters. Based on this facts, we can conclude that the age of HIP 66072 likely falls in a range of 40 to 125 Myr. Only Binks et al. (2015) have reported a stellar age, 27 Myr using evolutionary isochrones. HIP 66072 is listed as the suspected variable star, NSV 19865, showing an amplitude of 0.06-0.09 mag in \emph{V}-band. It has been classified as a rotational variable with a period of approximately 1.04 days. This rotational period was confirmed by Blinks (private communication) using TESS photometric data. While the gyrochronology technique is generally unsuitable for very young and extremely fast-rotating stars within the C-sequence (Mamajek \& Hillenbrand 2008), it may suggest that HIP 66072 is indeed a very young star. 
 
Kervella, Arenou \& Th\'evenin (2022), using Hipparcos and Gaia data, detected a significative proper motion anomaly. HDS 1898 has a semimajor axis of approximately 11 AU, and in 2016 the physical separation of stellar components was 7.6 AU. Therefore, the binarity of HDS 1898 could be the cause of the anomaly. \\

\underline{WDS 14087+3341 = HU 742 = ADS 9126} \\

HU 742 ( = HIP 69102) consists of stars with \emph{V} = 9.0 and 10-12 mag, approximately. It was initially observed by Hussey (1904) and, since 1995, the observations of the secondary star have covered more than 90 deg. The astrometric points show significant residuals with respect  to the last orbital solution (Popovich 1972). Our orbital solution have been previously announced in the IAUDS No 178 (Rica 2012b).

Our results suggests that the SpT of the primary star is G4-5V. Although, we find in the literature a SpT K0III. It is clear that the brightest component is not evolved because has a \emph{M}$_v$ = +4.7. Moreover, this system is catalogued as a suspected variable star, NSV6567, and is included in Baize \& Petit (1989) catalogue of double stars with variable components. The ASAS-SN catalogue (Jayasinghe et al. 2018) classified it as a type variable GCAS: characterized by eruptive irregular variations similar to the stars of $\gamma$ Cas type, with a period of 336 days and a photometric amplitude of 0.4 mag. In contrast, the AAVSO International Variable Star Index VSX (Watson et al. 2006-2014) reports a photometric variability in the range of \emph{V} = 9.5 – 12 mag. According to the historical observations of the WDS catalogue, the secondary component is probably  the variable star.

Gaia DR3 registers this object with a parallax very similar to that of Hipparcos. Gaia did not resolve the binary system, and the RUWE parameter is extremely high, 18.66, possibly due to the presence of the secondary star. \\

\underline{WDS 17130+0745 = STT 325 = ADS 10398}  \\

STT 325 (also known as HIP 84230 or HD 155714) is a binary star system comprising two stars with \emph{V} = 6.9 and 8.6 mag, separated by 0.35” and located at a distance of 95.6 pc. It was first observed in 1843 by Maedler (1844). We present the first orbital solution for STT 325, with a preliminary orbital period of 2828 years. Our orbital solution was previously  announced in the IAUDS No 178 (Rica 2012b).  Since the discovery of STT 325, the angular separation between the stars has decreased from 1.7” to 0.35” and the position angle has increased about 91 deg. 

According to the literature, STT 325 is a metal rich system with a combined SpT of F0, showing a [Fe/H] that ranges from +0.21 to +0.35 (Marsakov \& Shevelev 1995; Schkov et al. 2003; Anderson \& Francis 2012). The A component is a SB1 spectroscopic binary with a circular orbit as determined by Carquillat et al. (2004),  having an orbital period of 3.33 days.

Our analysis, assuming a [Fe/H] = +0.2 and with our estimated extinction of \emph{A}$_v$ = 0.15 mag, indicates that this system is composed of an A7V+G9V SB1 binary star and a F6V single secondary component, with masses of 1.77, 0.95 and 1.30 $M_\odot$, summing a total mass of about $4.0 \pm 0.2$ $M_\odot$. The isochrone that best matches the observational photometry is that corresponding to 0.9 Gyr old, in excellent agreement with the Geneva-Copenhagen Survey of Solar Neighbourhood (Holmberg 2007), which provides an age of 1.1 Gyr.

On the other hand, in agreement with the age determined from the galactocentric velocity by Holmberg et al. (2009), (\emph{U},\emph{V},\emph{W}) = (44.4, -7.6, -8.6) km s$^{-1}$, the stellar system matches with the thin disk population with young-medium age.

In order to derive individual properties for the stars within the SB1 binary system, we apply the methodology of Docobo \& Andrade (2006). This technique allows us to determine a \emph{M}$_v$ = +2.0 mag for STT 325 Aa and a $\Delta$\emph{V} of $3.5 \pm 0.4$ mag. Using the absolute magnitudes of Aa and Ab components, we infer additional astrophysical properties (see Table 4). We trust on Hipparcos parallax more because Gaia DR3 provides an astrometric solution with a very high RUWE value, 23.00, which means that this system is potentially influenced by binarity. Finally, applying the expressions published by Abushattal, Docobo \& Campo (2020), we obtain for the SB1 binary, orbital parameters, for the first time: a = $0.64 \pm 0.04$ mas (corresponding to 0.06 UA) and an inclination i = $5.2 \pm 0.3$ deg. 

Our sum of the stellar masses ($4.0 \pm 0.2$ $M_\odot$) is consistent with that estimated from the orbital parameters and the Hipparcos data. \\

\underline{WDS 21115+2144 = COU 227} \\ 

COU 227 (HIP 104612 = BD+21 4489) constitutes a binary star system with two components showing a  \emph{V} =  10.0 and 11.6 mag, and separated by 0.51”. Discovered by Couteau (1968) in 1967, it has been measured 23 times. The astrometric data presented in our  prior publication (Paper I) exhibit substantial deviations when is compared to the ehemerids of the earlier orbit. Our preliminary orbital solution (see Figure 14) was previously announced in the IAUDS No 176 (Rica 2012a). Over the course of its observation history, the secondary component has covered about 110 deg. The visual measurements show a large dispersion, making the orbital parameters presented in this study very preliminary. Consequently, further observations are needed to seed light to trajectory of the secondary star. 

We classify the stellar components as K0/1V and K5V stars. The total mass obtained using our orbital solution is higher than the expected mass based on SpT. Gaia DR3 did not resolved this binary and provides an astrometric solution with a very high RUWE, 23.00, which suggests a possible impact due to the binarity nature of this system.

Tsantaki et al. (2022) reported a radial velocity, $-4.8 \pm 2.9$ km s$^{-1}$, calibrated using Gaia DR2 radial velocity data. We calculate the first galactocentric velocity for COU 227, (\emph{U},\emph{V},\emph{W}) = (+27.0, -12.5, +8.0) km s$^{-1}$,  which would agree with a medium age object in the thin galactic disk. \\

\underline{WDS 22045+1551 = BU 696 AB = ADS 15599} \\

This binary ( = HD 209622 = HIP 108961) consists of two stars with magnitudes \emph{V} = 8.5 and 8.8, separated by 0.19” at a distance of 93 pc. The first measurement of BU 696 AB was registered in 1877.72 by Burnham but published in 1879 (Burnham 1879). However, the discovery measure was performed by Dembowsky (1883) in 1877.32. Subsequently, it has been observed on 53 additional times. We conducted measurements of this binary system in Rica et al. (2022). Since 1995, the secondary star shows a very important change in its position, already passing the apparent apoastro. We derived, for the first time, a highly inclined orbit. The historical measures cover a nearly complete revolution. Our orbital elements have previously been announced in the IAUDS No 180 (Rica 2013b). 

The differential magnitude from Hipparcos ($\Delta$\emph{V} = $1.68 \pm 0.75$ mag) is wrong. From visual photometric data compiled in the WDS catalogue, we estimate a $\Delta$\emph{V} = $0.3 \pm 0.1$ mag. This is the value we use in our study. We find in the literature references citting combined SpT of G5 (ppN catalogue), G0V (Abt, Corbally \& Christopher 2000) and G0IV (Stephenson \& Sanwal 1969). This system presents a rich metal content, [Fe/H] = +0.32 (Casagrande et al. 2011) and +0.34 (Brewer \& Fischer 2018).

We use a PARSEC isochorone with [Fe/H] = +0.30 and \emph{A}$_v$ = 0.05 to decompose the combined \emph{BVIJHK} photometry into individual fundamental parameters. Our findings reveal that this binary system comprises of F6V+F7V stars, with masses of 1.30 and 1.25 $M_\odot$, and an estimated age about 2.5 Gyr, which is in agreement with other values present in the literature, 2.4 - 2.5 Gyr (Casagrande et al. 2011; Mints \& Hekker 2017).

The dynamical parallax and masses that we obtain here agree with the Hipparcos trigonometric parallax and the expected total mass. Additionally, Casagrande et al. (2011) previously determined a galactocentric velocity (\emph{U},\emph{V},\emph{W}) of (+19, -40, +3) km s$^{-1}$. This velocity corresponds to a star within the old stellar population of the thin Galactic disk. \\

\underline{WDS 22052+2952  = A 893 = ADS 15610} \\ 

A 893, also identified as HD 209745 or HIP 109018, was discovered by Aitken (1904). This binary system comprises stars with \emph{V} = 9.0 and 10.0 mag, situated at a distance of 129 pc. Since 1904, it has been observed 29 times, the latest being in 2013. From seven measurements presented in Paper I, performed between 2010 and 2013 and additional historical astrometric points, we obtain a new orbital solution. The computed RMS for our eight measures is 1.0 deg ( = 4 mas) for $\theta$, and 9 mas for $\rho$. Our orbital elements have previously been announced in the IAUDS No 182 (Rica 2014b). The dynamical parallax agrees well with the Hipparcos trigonometric one.  

All measurements cover slightly more than one orbital revolution. The preceding orbital solution, published by Baize (1986), shows important residuals. Our new orbit is less eccentric and has a shorter period respect to the Baize’s orbit. In the literature we find combined SpT of F7V (Clausen \& Jensen 1979) and F8V (Wilson \& Joy 1950; Heard 1956).

We find that this stellar system is composed of F5V and G1V stars with a estimated age of $1.4_{-0.6}^{+1.0}$ Gyr (at 95\% confidence level in the interval [0.4 , 2.8] Gyr) and a [Fe/H] = +0.1 (at 95\% confidence level of [-0.10, +0.21]).  These values are in good agreement with the result of Ammons et al. (2006) that reported a [Fe/H] = +0.15 ± 0.13, and marginally agree with that reported in the StarHorse2 catalogue, [Fe/H] = $-0.15_{-0.09}^{+0.22}$.

It is noteworthy that Gaia DR3 lists this system with a significant error in parallax, $\pi$= $4.75 \pm 0.53$ mas, and proper motion. The high RUWE, 23.47, strongly suggests that the Gaia single star astrometric model is likely affected by the binary nature of HD 209745. Bobylev et al. (2006) determined a galactocentric velocity (\emph{U},\emph{V},\emph{W}) of (-13.1, -26.6, -12.9) km s$^{-1}$ which, in this work, we determine that corresponds to a star within the young stellar population in the thin Galactic disk. However, this kinematic motion, has not been associated with any known young kinematic group.

\section{Conclusions}

The measures presented in Paper I,  carried out with the CST telescope and the lucky-imaging FastCam camera, allowed us to determine or refine the orbits for the 15 objects analysed in detail in this work. By using two different orbital computational methods, we improved orbital parameters for 11 binaries (AG 14, D 5 AB, A 1581, HO 525 AB, WOR 19, A 1999, HU 572, HU 742, COU 227, BU 696 AB and A 893) and presented, for the first time, the orbital parameters for BU 1292, STF 147, HDS 1898 and STT 325. 

In order to determine individual photometric and fundamental properties for the binary components, we deblended the combined observational multiband photometry using PARSEC isochrones and our tool ``Binary Deblending'' (version 5.0). We found a binary composed of M1-2V dwarfs (WOR 19), a binary of evolved components (BU 1292 with two stars of F6IV-V) with a wide (10.5”) and a red (M3-4V) and faint (\emph{G} = 17.05 mag) companion reported for the first time. We also found an X-ray emissor (STF 147) and a very young quadruple system, WDS 04573+5345. 

The binaries here studied allowed us to determine their individual stellar masses, which are in agreement with the dynamical masses and, in some cases, the fundamental parameters of all members of the systems. The methodology exposed in this work even allows us to investigate systems composed by three and four components, in some cases not visually resolved. In general, the properties and features of multiple binaries here reported, will be used in future works in order to clarify the formation and evolution of these systems.

\begin{acknowledgement}
Rafael Barrena acknowledges support by the Severo Ochoa 2020 research programme of the Instituto de Astrof\'isica de Canarias. This article is based on observations made with the 1.52-m Carlos S\'anchez Telescope operated on the island of Tenerife by the Instituto de Astrof\'isica de Canarias in the spanish Observatorio del Teide. This research has made use of the Washington Double Star Catalog maintained at the U.S. Naval Observatory. 
\end{acknowledgement}

\end{document}